\documentstyle[12pt]{article}
 \newtheorem{defi}{Definition}
 \newtheorem{lem}{Lemma}
 \newtheorem{thm}{Theorem}
 \newtheorem{prop}{Proposition}
 \newtheorem{cor}{Corollary}
 
 \def\tin{|\!\!\!\!\bigcup}

 \def\det{\mathop{\rm det}}
 \def\tr{\mathop{\rm tr}\nolimits}
 \def\id{\mathop{\rm Id}\nolimits}
\def\Int{\mathop{\rm Int}\nolimits}
\def\Cup{\mathop{\cup}} 

\def\trh{\tr_{{\cal H}}}
\def\trt{\tr_{T_{\rho}P}}
\def\intt{\int_{T_{\rho}P}}

\def\lll{\langle }
\def\ggg{\rangle }

\def\idt{\id_{T_{\rho}P}}
\def\idh{\id_{{\cal H}}}
 \def\trp{T_{\rho}P}
 \def\utrt{{\cal U}(\trp)}
 \def\im{\mathop{\rm Im}\nolimits}
 \def\Hom{\mathop{\rm Hom}\nolimits}
 \def\End{\mathop{\rm End}\nolimits}
 \def\Ker{\mathop{\rm Ker}\nolimits}

\def\Spur{\mathop{\rm Spur}\nolimits}
\def\tl{{\cal T}_{sa}({\cal H})}
\def\trp{T_{\rho}P}
\def\trpd{T_{\rho}^{*}P}
\def\trpa{T_{\rho}P_{1}}
\def\trpb{T_{\rho}P_{2}}

\def\trpas{T_{\rho}^{*}P_{1}}

\def\trpbs{T_{\rho}^{*}P_{2}}

\def\aki{\vspace{3ex}}
\def\aku{\vspace{2ex}}
\def\ak{\vspace{1ex}}
\def\crb{Cram\'{e}r-Rao type bound}

\makeatletter
\newenvironment{pf}{\ak\noindent{\bf Proof}\quad}{\leavevmode\hfill$\Box$\par\@endpetrue}
\makeatother
\def\Label#1{\label{#1}\ [\ #1\ ]\ }
\def\det{\mathop{\rm det}}
\def\tr{\mathop{\rm tr}\nolimits}

\def\re{\mathop{\bf R}\nolimits}

\def\crb{Cram\'{e}r-Rao type bound}
\def\cri{Cram\'{e}r-Rao inequality}
\let\Label=\label
\makeatletter
\expandafter\def\csname eqnarray+\endcsname{\let\\=\@Eqncr \tabskip\@centering
   $$\halign to \displaywidth\bgroup\@eqnsel\hskip\@centering
   $\displaystyle\tabskip\z@{##}$&\hfil${{}##{}}$\hfil
   &$\displaystyle\tabskip\z@{##}$\hfil\tabskip\@centering
   &\hbox to \z@{\hss{##}}\tabskip\z@\cr}
\expandafter\def\csname endeqnarray+\endcsname{\crcr\egroup$$}
\def\@Eqncr{\cr\noalign{\penalty\interdisplaylinepenalty\vskip\jot\relax}}
\makeatother
\begin{document}
\vskip 2em \begin{center}
 {\LARGE\bf 
A Linear Programming Approach \par
to Attainable Cram\'{e}r-Rao type Bounds \par
and Randomness Condition
\par} \vskip 1.5em 
\large \lineskip .5em
Masahito Hayashi \par
Laboratory for Mathematical Neuroscience, 
Brain Science Institute, RIKEN,
2-1 Hirosawa, Wako, Saitama, 351-0198, Japan \par
e-mail address: masahito@brain.riken.go.jp
\end{center}
\begin{abstract}
The author studies the Cram\'{e}r-Rao type bound by a linear programming approach.
By this approach, he found a necessary and sufficient condition that
the Cram\'{e}r-Rao type bound is attained by a random measurement.
In a spin 1/2 system, this condition is satisfied.
\end{abstract}

\section{Introduction}
It is well-known that the lower bound of 
quantum \cri $~V_{\rho}(M) \ge J_{\rho}^{S}$ cannot be attained 
unless all the SLDs commute, where 
we denote by $V_{\rho}(M)$
a covariance matrix for a state $\rho$ by a measurement $M$,
the SLD Fisher information matrix for a state $\rho$ by $J_{\rho}^{S}$.
We therefore often treat an optimization problem for $\tr gV_{\rho}(M)$ 
to be minimized, where $g$ is an arbitrary real positive symmetric matrix. 
If there is a function $C_{\rho}$ (possibly depending on $g$) such that 
$\tr g V_{\rho} \ge C_{\rho}$ holds for all $M$, 
$C_{\rho}$ is called a \crb, or simply a CR bound. 
Our purpose is to find the most informative (i.e. attainable) CR bound 
under locally unbiasedness conditions. 

There is a few model, in which the attainable Cramer-Rao type bound
is calculated.
To author's knowledge,
 there has been known only two mixed state models for which 
this optimization problem was explicitly solved.
One is the estimation of complex amplitudes of coherent signals 
in Gaussian noise solved by Yuen and Lax, and Holevo. 
See Ref. 1. 2. 
Another one is the estimation of a 2-parameter spin 1/2 model solved 
by Nagaoka. See Ref. 3. 
Otherwise, recently pure state models have been studied on advanced level by 
Matsumoto, Fujiwara and Nagaoka. See Ref. 10. 4.

In two parameter case, Fujiwara and Nagaoka study 
the minimization problem for $\tr g V_{\rho}(M)$
in random measurements. See Ref. 10.
In this paper, in multi-parameter case, this minimization problem is explicitly solved in \S \ref{random}.

Otherwise, 
in \S \ref{secti2} and Appendix \ref{sec3}, a technique to calculate the set
$\{ V_{\rho}(M) | M $ is locally unbiased measurement at $\rho \}$ 
is introduced.

In \S \ref{soutui}, a completely different approach to the optimization problem 
is given based on an infinite dimensional linear programming technique.
See Ref 5.
By this approach, the minimization problem is translated 
into the other maximization problem.
In finite dimensional case, this maximization problem has the maximumvalue.  

In \S \ref{randomness}, we drive a necessary and sufficient condition that
the optimal measurement in \S \ref{random} is the optimal under the
locally unbiasedness conditions.
This condition is called the randomness condition.

In \S \ref{quadra}, it is proved that
when the dimension of quantum system is 2, any model satisfies the condition.

In general, $\lll~,~ \ggg$ denotes the linear pairing between a linear space 
and the dual.
$\langle ~|~ \rangle$ means a inner product on a linear space. 

\section{SLD inner product and locally unbiased conditions}\Label{sec1}
For $\rho \in {\cal T}_{sa}^{+}({\cal H})$, the space $L_{sa}^{2}(\rho)$ is defined as follows, where
${\cal T}_{sa}^{+}({\cal H})$ denotes the set of
$\{ \rho \in {\cal T}_{sa}({\cal H})| \rho \ge 0 \}$. 
\begin{defi}\quad
For $\rho \in {\cal T}_{sa}({\cal H})$, $L_{sa}^{2}(\rho)$ consists of
selfadjoint operators $X$ on ${\cal H}$ satisfying the following
conditions:
\begin{eqnarray}
&\circ& \phi_{j} \in {\cal D}(X) \hbox{ with respect to }j\hbox{ such that }
s_{j} \neq 0 \\
&\circ& \langle X | X \rangle^{sa}_{\rho} := \sum_{j}s_{j} \langle X \phi_{j}  |X \phi_{j} \rangle \,< \infty, 
\end{eqnarray}
where $\rho =\sum_{j}s_{j}|\phi_{j} \rangle \langle \phi_{j}| $ 
is the spectral decomposition of $\rho $. 
\end{defi}
For $X,Y \in L_{sa}^{2}(\rho)$, define:
\begin{eqnarray} 
\langle X |Y \rangle^{sa}_{\rho}:= \frac{1}{4} \Bigl (
\langle X + Y | X+Y \rangle^{sa}_{\rho} 
-\langle X - Y | X-Y \rangle^{sa}_{\rho} \Bigr ).
\end{eqnarray}
The inner product is called the SLD inner product, and
$\| ~~\|^{S}_{\rho}$ denotes the norm with respect to this inner product.
\begin{lem}\Label{quotient}
For $X \in L^2_{sa}(\rho)$, the following conditions are equivalent:
\begin{eqnarray}
&\circ & \langle X | X \rangle^{sa}_{\rho} = 0 \\
&\circ & X \rho + \rho X = 0 \\
&\circ & X \rho X = 0 \\
&\circ & \langle X | Y \rangle^{sa}_{\rho} = 0 ,  \hbox{ for }  \forall Y \in
L^2_{sa}(\rho) \\
&\circ & X \rho Y + Y \rho X = 0 ,  \hbox{ for }  \forall Y \in
L^2_{sa}(\rho) 
\end{eqnarray}
\end{lem}
${\cal L}_{sa}^2(\rho)$ denotes the quotient space $L_{sa}^2(\rho)/ K^2_{sa}(\rho)$.
From Lemma \ref{quotient}, for $X ,Y \in {\cal L}^2_{sa}(\rho)$
the following are independent of a lifting $T$ of the projection 
$L^2_{sa}(\rho) \to {\cal L}^2_{sa}(\rho)$:
\begin{eqnarray}
\langle X | Y \rangle_{\rho}^{sa} &:=& \langle T(X) | T(Y) \rangle_{\rho}^{sa} \\
\frac{1}{2}(X \rho Y+ Y \rho X) &:=& 
\frac{1}{2}(T(X) \rho T(Y) + T(Y) \rho T(X)) .
\end{eqnarray}

\begin{thm}\quad
If ${\cal H}$ is separable, ${\cal L}^{2}_{sa}(\rho )$ is a real Hilbert space with
 respect to the SLD inner product.
\end{thm}
For a proof see Ref. 6. 
\par\ak
We define $\rho \circ X :=\frac{1}{2} ( \rho \cdot X + X \cdot \rho ) 
\in {\cal T}_{sa}({\cal H})$ 
for $X \in {\cal L}^{2}_{sa}(\rho)$.
$J^{S}_{\rho}$ denotes
the inner product of the real Hilbert space ${\cal L}^{2}_{sa}(\rho)$.
${\cal L}^{2,*}_{sa}(\rho):=\{ \rho \circ X | X \in {\cal L}^{2}_{sa}(\rho) \}$
is regarded as the dual of ${\cal L}^{2}_{sa}(\rho)$ in the following: 
\begin{eqnarray*}
\begin{array}{ccc}
{\cal L}_{sa}^{2,*}(\rho) \times {\cal L}_{sa}^{2}(\rho) & \to & {\bf R} \\
\tin &  &\tin \\
(x,X) & \mapsto & \trh x X .
\end{array}
\end{eqnarray*}
$J^{S}_{\rho}$ can be regarded as an element of 
$\Hom_{sa}({\cal L}^{2}_{sa}(\rho),{\cal L}^{2,*}_{sa}(\rho))$ by
\begin{eqnarray}
\begin{array}{cccc}
J^{S}_{\rho} : & {\cal L}^{2}_{sa}(\rho) & \to & {\cal L}^{2,*}_{sa}(\rho) \\
& \tin & & \tin \\
& X & \mapsto & \rho \circ X .
\end{array}
\end{eqnarray}
\begin{defi}
\quad 
For a subset $\Theta \subset {\bf R}^{n}$ the map 
$f : \Theta \to {\cal T}_{sa}({\cal H})$ is called a $C^{k}$-map, if
the $k$-th derivative of $f$ is well defined on the interior of $\Theta$,
where ${\cal T}_{sa}({\cal H})$ is 
the set of selfadjoint trace class operators on ${\cal H}$.
\end{defi}
${\cal T}_{sa}^{+,1}({\cal H})$ denotes the set of
$\{ \rho \in {\cal T}_{sa}({\cal H})| \rho \ge 0 ,~ \trh \rho = 1 \}$. 
\begin{defi}\quad
We call $P \subset {\cal T}_{sa}^{+,1}({\cal H})$ an $n$-dimensional 
model, if there exist $\Theta \subset {\bf R}^n$ and $\phi : \Theta \to P$
such that $\phi$ is homeomorphism on the norm topology and $C^1$-map.
\end{defi}
In this paper,
$\frac{\partial}{\partial \theta^{i}}\in \trp$ is identified with 
$\frac{\partial \phi}{\partial \theta^{i}} \in {\cal T}_{sa}({\cal H})$. 
In this identification, we assume that $\trp$ is a subset of 
${\cal L}_{sa}^{2,*}(\rho)$.
For simplicity, we denote $J_{S}^{\rho}|_{\trpd}$ by $J_{S}^{\rho}$, too.
$\trpd$ is identified with $J_{S}^{\rho,-1}(\trp)$.
The inner product $J_{S}^{\rho,-1}$ on $\trp$ is called 
the *SLD inner product and $\| ~ \|_{S}$ denotes this norm.
In this paper, $n$ denotes the dimension of $\trp$.
${\cal M}(\Omega,{\cal H})$ denotes the set of generalized measurements
on ${\cal H}$ whose measurable space is $\Omega$.
For $X \in {\cal L}^2_{sa}(\rho)$, $M_X^T$ denotes the spectral decomposition 
of $T(X)$.
\begin{defi}\quad
An affine map $E$ from ${\cal M}(\trp , {\cal H})$ to 
$\Hom({\cal T}_{sa}({\cal H}),\trp)$ is defined by
\begin{eqnarray}
E(M)(\tau):= \intt x \trh \Bigl( M(\,d x) \tau \Bigr), ~
\forall \tau \in {\cal T}_{sa}({\cal H})
\Label{hom1}.
\end{eqnarray}
\end{defi}
Let us define the locally unbiasedness conditions.
\begin{defi}\quad
A measurement $M \in {\cal M}(T_{\rho}P,{\cal H})$ is called 
a locally unbiased measurement at $\rho \in P$, 
if the map $E(M) : {\cal T}_{sa}({\cal H}) \to T_{\rho} P $ 
satisfies the following conditions:
\begin{eqnarray}
E(M)(\rho) &=& 0 \Label{zero}\\
E(M)|_{T_{\rho}P} &=& \idt .
\Label{id}
\end{eqnarray}
${\cal U}(T_{\rho}P)$ denotes the set of locally unbiased measurements on $\rho \in P$.
\end{defi}
\begin{lem}\Label{localub}\quad
For $M \in {\cal M}(T_{\rho}P,{\cal H})$,
the condition (\ref{id}) is equivalent to the following equation:
\begin{equation}
\int_{T_{\rho}P} \tr_{{\cal H}} a(x) M(\,d x) =\tr_{T_{\rho}P} a 
,~ \forall a \in \End(T_{\rho}P) 
. \Label{tra}
\end{equation}
\end{lem}
By taking basis, it is easy to verify this.  
\par
Let $g $ be a nonnegative inner product on $\trp$,
then $\inf_{M \in \utrt} \trt V_{\rho}(M) g$ is called
the attainable Cram\'{e}r-Rao type bound,
 where $V_{\rho}(M) := \int_{\trp} x \otimes x \trh ( M(\,d x) \rho )$ 
is the covariance matrix. \par
Next, we consider locally unbiased and random measurements
(i.e. convex combinations of simple measurements).
$P(\trp \times \trpd)$ denotes the set of probability measures on 
$\trp \times \trpd$. The element $p$ of $P(\trp \times \trpd)$ is regarded 
a random measurement as:
\begin{eqnarray}
\begin{array}{cccc}
M_m^T : & P(\trp \times \trpd) & \to & {\cal M}(\trp,{\cal H}) \\
& \tin & & \tin \\
& p & \mapsto & \int_{\trp} \int_{\trpd} M^T(x,X) p(\,d x , \,d X)
\end{array}
\end{eqnarray}
where,
\begin{eqnarray*}
\begin{array}{cccc}
M^T: & \trp \times \trpd & \to & {\cal M}(\trp,{\cal H}) \\
& \tin & & \tin \\
& (x,X) & \mapsto & (M_X^T) \circ (x)^{-1} \\
&&&\\
(x) :& {\bf R} & \to & \trp \\
& \tin & & \tin \\
& c & \mapsto & c x . 
\end{array}
\end{eqnarray*}
Therefore, the set ${\cal U}_R(\trp) := P(\trp \times \trpd) \cap 
M_m^{T-1}({\cal U}(\trp))$ is regarded the set of locally unbiased and random measurements. 
The set ${\cal U}_R(\trp)$ is independent of $T$.
\begin{lem}
For $p \in P(\trp \times \trpd)$, $p$ is a locally unbiased measurement iff
\begin{eqnarray}
\int_{\trp} \int_{\trpd} \lll X , a(x) \ggg p(\,d x , \,d X) = \trt a, 
\forall a \in \End(\trp).
\end{eqnarray}
\end{lem}
It is trivial from Lemma \ref{localub}.
\begin{lem}
For $p \in P(\trp \times \trpd)$, the covariance matrix of $p$ is described as follows:
\begin{eqnarray}
V_{\rho} \circ M_m^T (p) = 
\int_{\trp} \int_{\trpd} \| X \|^2 x \otimes x ~p(\,d x , \,d X) . \Label{random2} 
\end{eqnarray}
\end{lem}
Since $V_{\rho} \circ M_m^T$ is independent of $T$,
$V_{\rho,R}$ denotes $V_{\rho} \circ M_m^T$.
\begin{defi}
We define the sets of covariance matrices in the following:
\begin{eqnarray*}
{\cal V}_{\rho} &:=&
\Bigl\{ V_{\rho}(M) \in S^+(\trp \otimes \trp)
\Bigl| M \in {\cal U}(\trp) \Bigr\} \\
{\cal V}_{\rho,R} &:=&
\Bigl\{ V_{\rho,R}(p) \in S^+(\trp \otimes \trp)
\Bigl| p \in {\cal U}_{R}(\trp) \Bigr\},
\end{eqnarray*}
where
$S(\trp \otimes \trp)$ denotes the symmetric tensor space 
of $\trp \otimes \trp$.
$S^+(\trp \otimes \trp)$ denotes the set of nonnegative elements of 
$S(\trp \otimes \trp)$.
\end{defi}
\begin{lem}\Label{affi}
${\cal V}_{\rho}$ and ${\cal V}_{\rho,R}$ are convex sets.
\end{lem}
\begin{pf}
${\cal U}(\trp)$ and ${\cal U}_R(\trp)$ are convex sets.
$V_{\rho} $ and $ M_m^T$ are affine maps.
Then ${\cal V}_{\rho}$ and ${\cal V}_{\rho,R}$ are convex sets.
\end{pf}

\section{Covariance matrix}\Label{secti2}
In this section we characterize ${\cal V}_{\rho}$ and 
${\cal V}_{\rho,R}$
For this purpose we need some definitions.
Let $W$ be a finite dimensional vector space.
We call a closed convex cone $L$ of $W$ a normal convex cone, if it satisfies the following conditions:
\begin{eqnarray}
&\circ& x \neq 0 \in L~,~\lambda \,< 0 \hbox{ }\Rightarrow \hbox{ }\lambda x \notin L \nonumber \\
&\circ& W=L+(-L). \Label{joukencov2}
\end{eqnarray}
Now we let $L$ a normal convex cone.
Let $g$ be an inner product such that satisfies the following condition:
\begin{eqnarray}
l_1,l_2 \in L~,~g(l_1, l_1) \ge  g(l_1 + l_2 ,l_1+l_2) \Longrightarrow l_2=0. \Label{joukencov} 
\end{eqnarray}
When $W$ is $S(\trp \otimes \trp)$, $S^+(\trp \otimes \trp)$ is 
a normal positive cone.
\begin{defi}\rm
A subset $C$ of $L$ is called {\it $L$-stable} set if 
\begin{eqnarray}
C = C + L.
\end{eqnarray}
\end{defi}
\begin{prop}\Label{lc1}
${\cal V}_{\rho}$ and ${\cal V}_{\rho,R}$ are
 $S^{+}(\trp \otimes \trp)$-stable and convex.
\end{prop}
\begin{pf}
From Lemma \ref{affi}, they are convex.
First we prove that ${\cal V}_{\rho}$ is $S^{+}(\trp \otimes \trp)$-stable. 
It is sufficient to show that 
$V_{\rho}(M) + x \otimes x \in {\cal V}_{\rho}$ for any 
$M \in \utrt $,and $x \in \trt $.
We define an affine map $S_{x}$ in the following way:
\begin{eqnarray}
\begin{array}{cccc}
S_{x}:& \trp & \to & \trp \\
& \tin & & \tin \\
& y & \mapsto & y + x
\end{array}
\end{eqnarray}
Let the map $M_{x}:=1/2(M \circ S_{x} + M \circ S_{-x})$. As $E_{M_{x}}=1/2((S_{x})^{-1} + (S_{-x})^{-1} )$,
 $M_{x} \in \utrt$.
\begin{eqnarray}
V_{\rho}(M_{x}) 
&=& \frac{1}{2} V_{\rho}(M \circ S_{x} )+\frac{1}{2} (M \circ S_{-x} ) 
\nonumber \\
&=& \frac{1}{2} \intt (y - x ) \otimes (y -x) + ( y + x )\otimes (y+x) \trh (M(\,d y) \rho) 
\nonumber \\
&=& \intt y \otimes y + x \otimes x \trh (M(\,d y) \rho) \nonumber \\
&=& V_{\rho}(M) + x \otimes x.
\end{eqnarray}
We obtain $V_{\rho}(M) + x \otimes x \in {\cal V}_{\rho}$.
Similarly it is proved that 
${\cal V}_{\rho,R}$ is $S^{+}(\trp \otimes \trp)$-stable. 
\end{pf}
From the quantum Cram\'{e}r-Rao inequality,
we get the following relation:
\begin{eqnarray}
{\cal V}_{\rho,R} \subset {\cal V}_{\rho} \subset \{ J^S_{\rho} \}.
\end{eqnarray}
To characterize a $L$-stable set $C$, we define the following set $K(C)$.
\begin{defi}\rm
For a subset $C$ of $L$, the {\it limit set} $K(C)$ of $C$ 
is defined as follows:
\begin{eqnarray}
K(C) :=\{ x \in C | (x - L) \cap C = \{ x \} \}.
\end{eqnarray}
\end{defi}
\begin{lem}
When a subset $C$ of $L$ is $L$-stable and closed, then 
$C= K(C) + L$.
\end{lem}
\begin{pf}
It suffices to verify that there exists an element $y \in (C)$ such that
$x \in y +L$ for arbitrary $x \in C \setminus K(C)$.
$(x - L) \cap C \subset L$ is a compact set.
Therefore, there exists $y \in (x - L) \cap C$ so that 
$g(z, z ) \ge f(y, y)$ for arbitrary $z \in (x - L) \cap C$.
Now we prove that $(y - L) \cap C = \{ y \}$ by reductive absurdity.
Let $z \in (y-L) \cap C~,~z \neq y$, then 
there exists $l \in L~,l\neq 0$ so that $y=z+l$.
Because $z,l \in L~,~l \neq 0$, from (\ref{joukencov}) 
\begin{eqnarray}
g(y,y) \,> g(z,z) \Label{yz}.
\end{eqnarray}
(\ref{yz}) contradicts the definition of $y$.
Hence $(y -L ) \cap C= \{ y \}$, thus $y \in K(C)$. 
Because $y  \in ( x - L) \cap C $, we conclude $x \in y + L$.
\end{pf}
In Appendix \ref{sec3}, we prove a useful theorem to calculate
$K({\cal V}_{\rho,R})$ and $K({\cal V}_{\rho})$.

\section{Random Limit}\Label{random}
Next, we minimize the following value ${\cal D}^{\rho}_{g,R}$ 
in locally unbiased and random measurements ${\cal U}_{R}(\trp)$.
\begin{defi}\rm\quad
The {\it deviation} ${\cal D}^{\rho}_{g,R}$ 
for a measurement $p \in P(\trp \times \trpd)$ is defined as follows:
\begin{eqnarray}
{\cal D}^{\rho}_{g,R}(p) 
:= \tr_{\trpd} g V_{\rho,R} (p) 
= \int_{\trp} \int_{\trpd} g(x,x) \| X \|^2 p(\, d X \,d x). \Label{sec2}
\end{eqnarray}
\end{defi}
We introduce the useful theorem to minimize the deviation 
${\cal D}_{g,R}^{\rho}(M)$ under the locally unbiasedness conditions.
\begin{thm}\quad\Label{mainr}
We have the inequality:
\begin{eqnarray}
 \inf_{M \in {\cal U}_R(T_{\rho}P)}{\cal D}^{\rho}_{g,R}(M) 
\ge \sup_{ (a,S) \in {\cal U}^{*}_R(g) } ( \trt a + S ) ,
\end{eqnarray}
 where 
\begin{eqnarray*}
{\cal U}^{*}_R(g) 
&:=&  \{ (a,S) \in \End(T_{\rho}P) \times {\bf R}  |
R_{g,R}^{\rho}(a,S;x,X) \ge 0
,~ \forall (x,X) \in \trp \times \trpd
\} \\
R_{g,R}^{\rho}(a,S;x,X) &:=& g (x , x ) \| X \|^2 - \lll X ,a(x) \ggg - S .
\end{eqnarray*}
\end{thm}
\begin{cor}\quad
\Label{howtor}
If there exist a locally unbiased and random measurement 
$p' \in P(\trp \times \trpd)$
and an element $(a',S')$ of ${\cal U}^{*}_R(g)$ satisfying the condition:
\begin{equation}
{\cal R}_{g,R}^{\rho}(a',S';p') = 0 \Label{hanteir},
\end{equation}
then we obtain 
\begin{eqnarray}
{\cal D}_{g,R}^{\rho}(p') = \trt a' + S'
 = \inf_{p \in {\cal U}_R(T_{\rho}P)}{\cal D}^{\rho}_{g,R}(p)=
\sup_{ (a,S) \in {\cal U}^{*}_R(g) } \trt a +S , 
\end{eqnarray}
where ${\cal R}_{g,R}^{\rho}$ is defined as:
\begin{eqnarray}
{\cal R}_{g}^{\rho}(a,S;p) :=
\int_{\trp} \int_{\trpd} R_{g,R}^{\rho}(a,S;x,X) p(\,d X \,dx ). 
\end{eqnarray}
\end{cor}
$(a,S) \in {\cal U}^{*}(g)$ is called the Lagrange multiplier.
\par\noindent
{\bf Proof of Theorem \ref{mainr} and Corollary \ref{howtor} }\quad
For $p \in {\cal U}_R(T_{\rho}P)$ and $(a,S) \in {\cal U}^{*}_R(g)$, we have
\begin{eqnarray}
&~& {\cal R}_{g,R}^{\rho}(a,S;p) \nonumber \\
&=& \int_{\trp} \int_{\trpd}  g( x ,x ) \| X \|^2 p(\,d X \,d x) 
-\int_{\trp} \int_{\trpd}  \lll X , a(x) \ggg p(\,d X \,d x) 
-\int_{\trp} \int_{\trpd}  S p(\,d X \,d x) \nonumber \\
&=& {\cal D}_{g,R}^{\rho}(p) - \trt a - S \Label{maincc}.
\end{eqnarray}
Since we have $R_{g,R}^{\rho}(a,S;x,X) \ge 0$ for 
$\forall (x,X) \in \trp \times \trpd$,
 we obtain ${\cal R}_{g,R}^{\rho}(a,S;p) \ge 0$.
 By (\ref{maincc}), the proof of Theorem \ref{mainr} is complete.
Substitute $(a,S)=(a',S'),p=p'$, then the proof of Corollary \ref{howtor} is complete. 
\leavevmode\hfill$\Box$\par
\begin{thm}
If $g=W^* J W,~ \trt W =1$, then 
\begin{eqnarray}
\inf_{p \in {\cal U}_R(\trp)}{\cal D}^{\rho}_{g,R}(p)
= 1 . \Label{random1}
\end{eqnarray}
The Optimal measurement is given by (\ref{ropt1}).
\end{thm}
\begin{pf}
Let Lagrange multiplier $(a,S)$ be $(2W,-1)$, then
\begin{eqnarray}
{\cal R}_{g,R}^{\rho}(2W,-1;x,X) =
\| W(x) \|^2 \| X \|^2 - 2 \lll X , W(x) \ggg + 1 \ge 0.
\end{eqnarray}
Let $W_i$ be an eigen value of $W$ and, $e_i$ be an eigenvector of $W$, where 
$\| e_i \| =1$.
$M^T_W$ is defined as follows:
\begin{eqnarray}
M^T_W:= \sum_{i=1}^n W_i M^T(W_i ^{-1} e_i, J^{-1} e_i ) . \Label{ropt1}
\end{eqnarray}
Then $M^T_W \in {\cal U}_R(\trp)$ and,
\begin{eqnarray}
{\cal R}_{g,R}^{\rho}(2W,-1;M^T_W)
= \sum_{i=1}^n W_i (\| e_i \|^2 \| J^{-1}e_i \|^2 
- 2 \lll J^{-1}e_i , e_i \ggg + 1 )
=0.
\end{eqnarray}
$M^T_W$ and $(2W,-1)$ satisfy the condition of Corollary \ref{howtor}.
 Since $\trt 2W -1 = 1$, we obtain (\ref{random1}).
\end{pf}
When a state $\rho$ is measured by the measurement $M^T_W$,
 the following covariance matrix by (\ref{random2}).
\begin{eqnarray}
V_{\rho}(M^T_W) 
= \sum_{i=1}^n W_i (W_i^{-1} e_i) \otimes  (W_i^{-1} e_i) \| e_i\|^2 
= \sum_{i=1}^n W_i^{-1} e_i \otimes  e_i 
= W^{-1} J .
\end{eqnarray}
From the preceding proof, the map $Q_R$ is derived in the following:
\begin{eqnarray}
\begin{array}{cccc}
Q_R:& S(\trpd \otimes \trpd )& \to & S(\trp \otimes \trp ) \\
& \tin & & \tin \\
& W^* J W & \mapsto & \frac{W^{-1}J}{\trt W} .
\end{array} 
\end{eqnarray}
$\im Q_R=\{ W^{-1} J | \trt W = 1 \}$ is closed.
Since this map is $S^+(\trp \otimes \trp)$-conic, we have the following Theorem.  
\begin{thm}
The limit set of ${\cal V}_{\rho,R}$ is described as follows:
\begin{eqnarray}
K({\cal V}_{\rho,R}) = 
\{ W^{-1} J | \trt W = 1 \} .
\end{eqnarray} 
This limit set is called the random limit.
\end{thm}
\begin{lem}
In two parameter case, the random limit is described below:
\begin{eqnarray}
K({\cal V}_{\rho,R}) = 
\{ J + X J | \det X =1 \}.
\end{eqnarray}
\end{lem}
\section{Linear programming approach}\Label{soutui}
We introduce a new approach to the attainable Cram\'{e}r-Rao type bound.
 In this approach, applying the duality theorem of the infinite dimensional 
linear programming, the bound is characterized.
 But, we don't have to know the duality theorem for this section.
If the reader is interested in the duality theorem, see Ref 5.
In the noncommutative case, 
there is no infimum of covariance matrices under the 
locally unbiasedness conditions.
Therefore, we minimize the following value ${\cal D}^{\rho}_{g}$ 
under the locally unbiasedness conditions.
Let $g$ be a nonnegative inner product on $\trp$.
\begin{defi}\quad
The deviation ${\cal D}^{\rho}_{g}$ 
for a measurement $M \in {\cal M}(\trp,{\cal H})$ is defined as follows:
\begin{eqnarray}
{\cal D}^{\rho}_{g}(M) 
:= \tr_{\trpd} g V_{\rho}(M) 
= \int_{\trp}g(x,x) \trh M(\,dx)\rho . \Label{deviation}
\end{eqnarray}
\end{defi}
Let us define a linear functional on $\End (\trp ) \times \tl $,
denoted by $\Spur$ in the following way.
We introduce a useful theorem to minimize the deviation 
${\cal D}_{g}^{\rho}(M)$ under the locally unbiasedness conditions.
\begin{thm}\quad\Label{main}
We have the inequality:
\begin{eqnarray}
 \inf_{M \in {\cal U}(T_{\rho}P)}{\cal D}^{\rho}_{g}(M) 
\ge \sup_{ (a,S) \in {\cal U}^{*}(g) } \Spur(a,S)  ,\Label{dua}
\end{eqnarray}
 where 
\begin{eqnarray*}
\Spur (a,S) &:=& \trt a + \trh S \\
{\cal U}^{*}(g) 
&:=& \{ (a,S) \in \End(T_{\rho}P) \times {\cal T}_{sa}({\cal H})  |
R_{g}^{\rho}(a,S;x) 
\ge 0 
,~ \forall x \in T_{\rho}P  
\} \\
R_{g}^{\rho}(a,S;x) &:=& g (x , x ) \cdot \rho - S - a(x) .
\end{eqnarray*}
Notice that $T_{\rho}P$ is a subset of ${\cal T}_{sa}({\cal H})$.
\end{thm}
The calculation of $\sup_{ (a,S) \in {\cal U}^{*}(g) } \Spur(a,S)  $ 
is called the dual problem.
\begin{cor}\quad
\Label{howto}
If there exist a sequence of locally unbiased measurements $\{ M_{k} \}$
and an element $(a',S')$ of ${\cal U}^{*}(g)$ satisfying the condition:
\begin{equation}
{\cal R}_{g}^{\rho}(a',S';M_{k}) \to 0 ~({\it as}~k \to 0) \Label{hantei},
\end{equation}
then
\begin{eqnarray}
\lim_{k \to \infty} {\cal D}_{g}^{\rho}(M_{k}) = \Spur(a',S')
 = \inf_{M \in {\cal U}(T_{\rho}P)}{\cal D}^{\rho}_{g}(M)=
\sup_{ (a,S) \in {\cal U}^{*}(g) } \Spur(a ,S ) , 
\end{eqnarray}
where ${\cal R}_{g}^{\rho}$ is defined as:
\begin{eqnarray}
{\cal R}_{g}^{\rho}(a,S;M) := \trh \int_{T_{\rho}P} R_{g}^{\rho}(a,S;x)M(\,dx).
\end{eqnarray}
\end{cor}
$(a,S) \in {\cal U}^{*}(g)$ is called the Lagrange multiplier.
\par\noindent
{\bf Proof of Theorem \ref{main} and Corollary \ref{howto} }\quad
For $M \in {\cal U}(T_{\rho}P)$ and $(a,S) \in {\cal U}^{*}(g)$, we have
\begin{eqnarray}
&~& {\cal R}_{g}^{\rho}(a,S;M) \nonumber \\
&=& \tr_{{\cal H}} \int_{T_{\rho}P}  g( x ,x ) \cdot \rho M(\,dx) - \tr_{{\cal H}} \int_{T_{\rho}P} S M(\,dx) - \tr_{{\cal H}} \int_{T_{\rho}P} a(x) M(\,dx)  \nonumber \\
g&=& {\cal D}_{g}^{\rho}(M) - \tr_{T_{\rho}P} a - \trh S \Label{mainc}.
\end{eqnarray}
Since $R_{g}^{\rho}(a,S;x) \ge 0$ for any $x \in T_{\rho}P$,
 we obtain ${\cal R}_{g}^{\rho}(a,S;M) \ge 0$.
 By (\ref{mainc}), the proof of Theorem \ref{main} is complete.
Substitute $(a,S)=(a',S')$, then the proof of Corollary \ref{howto} 
is complete.
\leavevmode\hfill$\Box$\par
\noindent
Indeed we obtain the following theorem.
A proof of Theorem \ref{mani} is too long.
See Appendix \ref{mainth}.
\begin{thm}\Label{mani}
We obtain 
\begin{eqnarray*}
 \inf_{M \in {\cal U}(T_{\rho}P)}{\cal D}^{\rho}_{g}(M) 
= \sup_{ (a,S) \in \tilde{{\cal U}}^{*}(g) } \Spur (a, S),
\end{eqnarray*}
where 
\begin{eqnarray*}
\tilde{{\cal U}}^{*}(g) 
&:=& \{ (a , S)\subset \End(T_{\rho}P) \times {\cal B}_{sa}^{*}({\cal H})  |
\forall x \in \trp ~,~R_g^{\rho}(a,S;x) 
\in {\cal B}_{sa}^{*,+}({\cal H}) \} \\
\Spur (a,S) &:=& 
\trt a + \lll S, \idh \ggg \\
R_g^{\rho}(a,S;x) &:=&
g(x,x) \cdot \rho - S - a(x). 
\end{eqnarray*}
 ${\cal B}_{sa}^{*}({\cal H})$ is the topological dual space of ${\cal B}_{sa}({\cal H})$ with respect to the norm topology.
By the preceding equation 
we have the equality in (\ref{dua}) in the case
of $\dim {\cal H} \,< \infty$.
But we don't know whether we have the equality in the case of
$\dim {\cal H} = \infty$.
The calculation of 
$\sup_{ (a,S) \in {\cal U}^{*}(g) } \Spur (a,S) $
is called the dual problem.
\end{thm}

\noindent
\subsection{Maximum}
In this section, we consider the dual problem.
$\trp$ is regarded as a real Hilbert space with respect to
 $J_{S}^{\rho,-1}$.
\begin{lem}\quad
If the dimension of ${\cal H}$ is finite,
 then the set ${\cal U}^{*}(g) \cap \Spur ^{-1}( [0, \infty) )$ is compact.
\end{lem}
 We assume that the norm of $\End (\trp )$ is the operator norm $\| ~\|_{o}$,
 and the norm of  $\tl$ is the trace norm $\| ~ \| _{t}$.
The norm $\|~\|_{o,t}$ of $\End(\trp) \times \tl$ is defined as follows:
\begin{eqnarray*}
\| (a,S) \|_{o,t}:= \| a \|_{o} + \| S \|_{t},
~\forall (a,S) \in \End(\trp) \times \tl .
\end{eqnarray*} 
\begin{pf}
We have
\begin{eqnarray*}
{\cal U}^{*}(g) = \cap_{x \in \trp} \{ (a,S) | g( x , x ) \cdot \rho - S - a(x) \in {\cal T}_{sa}^{+}({\cal H}) \} .
\end{eqnarray*}
Moreover, $\{ (a,S) | g( x , x ) \cdot \rho - S - a(x) \in {\cal T}_{sa}^{+}({\cal H}) \}$ is closed. 
 Thus, ${\cal U}^{*}(g) $ is closed.
 Because $\Spur^{-1}( [ 0, \infty) )$ is closed. 
 ${\cal U}^{*}(g) \cap \Spur ^{-1}( [0, \infty) )$ is closed.
 Therefore, it is sufficient to show
 that it is bounded with respect to the norm $\| ~ \|_{o,t}$.
 Denote $n := \dim \trp$.
For $( a,S) \in {\cal U}^{*}(g) \cap \Spur ^{-1}( [0, \infty) )$,
we have $\trt a \le n \| a \|_{o}$.
Choose $z \in \trp$ such that $\| z \|=1,~\|a(z)\|=\| a \|_{o}$.
 For $r \,> 0$, we have
\begin{eqnarray}
 g (r \cdot z ,r \cdot z ) \rho - a( r \cdot z ) - S \ge 0 . \Label{cpt1}
\end{eqnarray}
Substitute $r=0$, then $-S \ge 0$.
Let us calculate the left hand side of (\ref{cpt1}).
\begin{eqnarray}
&~& g (r \cdot z , r \cdot z ) \rho - a( r \cdot z ) - S 
= r^{2} \cdot g (z , z) \rho - r \cdot J^{\rho,-1}_{S}(a(z)) \circ \rho - S
\nonumber \\
&=& \Bigl( \sqrt{g( z , z)} r - \frac{1}{2 \sqrt{g( z , z )}} J^{\rho,-1}_{S}(a(z)) \Bigr)\cdot \rho \cdot \Bigl( \sqrt{g( z , z )} r - \frac{1}{2 \sqrt{g( z , z )}} J^{\rho,-1}_{S}(a(z)) \Bigr) \nonumber \\
&~& - \frac{1}{4 g( z , z )} J^{\rho,-1}_{S}(a(z)) \cdot  \rho \cdot J^{\rho,-1}_{S}(a(z)) - S .\Label{cpt8}
\end{eqnarray}
Let $\{ e_{i} \}$ be a complete orthonormal system of ${\cal H}$
which consists of eigenvectors of  $J^{\rho,-1}_{S}(a(z)) $.
Substitute $r$ for the eigenvalue $\alpha_{i}$ of
$\frac{1}{2 g( z , z)} J^{\rho,-1}_{S}(a(z)) $
corresponding to the eigenvector $e_{i}$, then we have
\begin{eqnarray*}
\Bigl\langle e_{i} \Bigl| \Bigl( \sqrt{g( z , z)} \alpha_i - \frac{1}{2 \sqrt{g( z , z )}} J^{\rho,-1}_{S}(a(z)) \Bigr)
\cdot \rho \cdot \Bigl( \sqrt{g( z , z )} \alpha_i - \frac{1}{2 \sqrt{g( z , z )}} J^{\rho,-1}_{S}(a(z)) \Bigr) \Bigr| e_i \Bigr\rangle = 0.
\end{eqnarray*}
By (\ref{cpt1}), we have
\begin{eqnarray*}
\Bigl\langle e_{i} \Bigl|  - \frac{1}{4   g( z , z )} J^{\rho,-1}_{S}(a(z)) \cdot  \rho \cdot J^{\rho,-1}_{S}(a(z)) - S \Bigr| e_{i} \Bigr\rangle \ge 0 .
\end{eqnarray*}
Sum up for $i$ from $1$ to $n$.
\begin{eqnarray*}
\trh \Bigl( - \frac{1}{4 g (z ,z )} J^{\rho,-1}_{S}(a(z)) \cdot \rho 
\cdot J^{\rho,-1}_{S}(a(z)) - S  \Bigr) \ge  0 .
\end{eqnarray*}
Thus, we get
\begin{eqnarray*}
\trh S \le - \frac{1}{4 g( z , z )} \langle a(z) | a(z) \rangle_{S}^{\rho} 
= - \frac{\| a \|_{o}^{2}}{4 g( z,z)}.
\end{eqnarray*}
Therefore, we obtain
\begin{eqnarray*}
0 \le  \Spur (a,S) \le n\| a \|_{o} - \frac{\| a \|_{o}^{2}}{4 \| g \|_{o}}. 
\end{eqnarray*}
Hence, $ 0 \le \| a \|_{o}(n - \frac{\| a \|_{o}}{4 \| g \|_{o}}) $.
 Thus, $0 \le \| a \|_{o} \le 4 n \| g \|_{o}$.
As $-S \ge 0$, we have
 $\| S \|_t = -\trh S $.
 Therefore, we obtain the following inequalities:
\begin{eqnarray*}
0 \le \| S \|_t \le \tr a \le n \| a \|_{o} \le 4 \| g \|_{o} n ^{2}.
\end{eqnarray*}
Thus, ${\cal U}^{*}(g) \cap \Spur ^{-1}( [0, \infty) )$ is bounded,
hence compact.
\end{pf}
We have the following corollary.
\begin{cor}\quad
There exists the maximum of the right hand side of (\ref{dua}).
\end{cor}
Assume that $\rho \in P_{1} \subset P_{2}$ and $\trpa \subset \trpb,~\trpa \neq \trpb$.
From the embedding map $i : P_{1} \hookrightarrow P_{2}$,
we have $\,d i_{\rho} : \trpa \hookrightarrow \trpb$ and 
$ \,d i_{\rho}^{*} : \trpbs \to \trpas$.
By identifying the dual $T_{\rho}^{*}P_{i}$ with $T_{\rho}P_{i}~(i=1,2)$,
$\,d i_{\rho}^{*}$ can be regarded
as $ \,d i_{\rho}^{*} : \trpb \to \trpa$.
 Let $g$ be a nonnegative inner product on $\trpa$,
then $\,d i_{\rho} g \,d i_{\rho}^{*} $ is a nonnegative inner product 
on $\trpb$.
\begin{lem}\quad\Label{sub1}
We have the inequality:
\begin{eqnarray}
\max_{(a,S) \in {\cal U}^{*}(g)}\Spur(a ,S ) \le
\max_{(a',S) \in {\cal U}^{*}(\,d i_{\rho} g \,d i_{\rho}^{*} )}
\Spur( a' , S ) .\Label{cpt2}
\end{eqnarray}
Moreover the equality in (\ref{cpt2}) holds, if and only if there exists $(a',S) \in {\cal U}^{*}(\,d i_{\rho} g \,d i_{\rho}^{*})$
such that $a' (\trpa ) \subset \trpa$, and the maximum of the right hand side
is attained by $(a',S)$.
\end{lem}
\begin{pf}
We have $(\,d i_{\rho} a \,d i_{\rho}^{*}, S ) \in {\cal U}^{*}(\,d i_{\rho} g \,d i_{\rho}^{*})$ 
for $(a,S) \in {\cal U}^{*}(g)$.
\begin{eqnarray}
\begin{array}{cccc}
F:& {\cal U}^{*}(g) & \to & {\cal U}^{*}(\,d i_{\rho} g \,d i_{\rho}^{*}) \\
& \tin & & \tin \\
& (a,S) & \mapsto & (\,d i_{\rho} a \,d i_{\rho}^{*} ,S).
\end{array}
\end{eqnarray}
Then, $\Spur(a,S)=\Spur(F(a,S))$.
Therefore we obtain Inequality (\ref{cpt2})
The equality holds in (\ref{cpt2}), if and only if
\begin{eqnarray}
\max_{(a',S) \in \im F} \Spur(a',S).
=\max_{(a',S) \in {\cal U}^{*}(\,d i_{\rho} g \,d i_{\rho}^{*})} \Spur(a',S) 
\end{eqnarray}
By the definition of ${\cal U}^{*}(\,d i_{\rho} g \,d i_{\rho}^{*})$,
 as $\,d i_{\rho} g \,d i_{\rho}^{*}(\Ker \,d i_{\rho}^{*})=0$,
we have $a'(\Ker \,d i_{\rho}^{*})=0$ for 
$(a',S) \in {\cal U}^{*}(\,d i_{\rho} g \,d i_{\rho}^{*})$.
Thus,
$(a',S) \in \im F$ for 
$(a',S) \in {\cal U}^{*}(\,d i_{\rho} g \,d i_{\rho}^{*})$,
if and only if $a'(\trpa) \subset \trpa$.
 Thus, the proof is complete.
\end{pf}
\section{Randomness condition}\Label{randomness}
\begin{thm}\Label{randomness1}
In the finite-dimensional case, the following four conditions are equivalent.
\begin{eqnarray*}
&(1)& \forall X,Y \in \trpd ,~\| X \| = \| Y \|=1  \Rightarrow 
X \rho X = Y \rho Y . \\
&(2)& \hbox{There exists a complete orthonarmal base} ~\{ X_1 , \ldots X_n \} 
~\hbox{of}~ \trpd \\ 
&~& \hbox{ such that } 
X_i \rho X_j + X_j \rho X_i = 0 ,~X_i \rho X_i = X_j \rho X_j 
~\hbox{ for }~(i \neq j) . \\
&(3)& {\cal V}_{\rho}= {\cal V}_{\rho,R} . \\
&(4)& \hbox{There exists}~ g \,> 0 \hbox{ such that } 
\inf_{M \in {\cal U}(\trp) } {\cal D}_g^{\rho} = (\trt \sqrt{J g})^2 . 
\end{eqnarray*}
\end{thm}
\begin{defi}\rm
If $\trp$ satisfies the preceding condition, $\trp$ is called 
a {\it random model}.
\end{defi}
\begin{pf}
$(3) \Rightarrow (4),~(2) \Leftrightarrow (1)$ is easy.
In this proof, $W_i$ denotes an eigenvalue of $W$, and 
$e_i$ denotes an eigenvector of $W$, where $\| e_i \|= 1 $. \par
Without loss of generality,
we can assume that $g= W^* J W,~W \in \End_{sa}(\trp),~\trt W =1 $.

For simplicity, $S(x)$ denotes $J^{-1}x \rho J^{-1}x$ for $x \in \trp$.
First, let's prove $(1) \Rightarrow (3)$.
For $g:=W^* J W$,
we calculate $\inf_{M \in {\cal U}(\trp)} {\cal D}_g^{\rho}$.
Take the Lagrange multipliers in the following way:
\begin{eqnarray*}
a &:=& 2 W \\
S &:=& - X \rho X ,
\end{eqnarray*}
where we put $X \in \trpd,~ \| X \| =1$.
 Then, we have
\begin{eqnarray*}
&~& R_{g}^{\rho}(2W,S ;y W^{-1} z)\\
&=& g( y W^{-1} z ,y W^{-1}z ) \cdot \rho - 2 W(y W^{-1} z) 
+ J^{-1}(z)\rho J^{-1}(z) \\
&=& y^2 \cdot \rho - 2 y z + J^{-1}(z)\rho J^{-1}(z) \\
&=& y^2 \cdot \rho - 2 y J^{-1}(z) \circ \rho + J^{-1}(z)\rho J^{-1}(z) \\
&=& \Bigl(y - J^{-1}(z) \Bigr)\rho \Bigl(y - J^{-1}(z) \Bigr) ,
\end{eqnarray*}
where $z \in \trp ,~ \| z \| =1,~ y \in {\bf R}$.
 Therefore, $(2W,S) \in {\cal U}^{*}(g)$.
 Thus, $\Spur( 2W , S) = 1 $
 is a \crb.
Substitute $M = M^T_W$, then
\begin{eqnarray*}
 {\cal R}_{g}^{\rho}(a,S;M^T_W)
= \sum_{i=1}^n W_i {\cal R}_{g}^{\rho}(a,S;M^T(W_i^{-1}e_i ,J^{-1}e_i)) 
= 0
\end{eqnarray*}
because
\begin{eqnarray*}
&~& {\cal R}_{g}^{\rho}(a,S;M^T(W_i^{-1}e_i ,J^{-1}e_i)) \\
&=& \trh \Bigl( \int_{{\bf R}} R_g^{\rho} (a,S; y W_i^{-1}e_i) M^T_{J^{-1}(e_i)} (\,d y) \Bigr) \\
&=& \trh \Bigl( \int_{{\bf R}} 
\Bigl(y - J^{-1}(e_i) \Bigr)\rho \Bigl(y - J^{-1}(e_i) \Bigr) 
M^T_{J^{-1}(e_i)} (\,d y) \Bigr) = 0 .
\end{eqnarray*}
As $(2W,S)$ and $M^T_W$ satisfy the conditions of Corollary \ref{howto},
 the random measurement 
$M^T_W$ attains a \crb $~1$.
Therefore $(1) \Rightarrow (3)$ is proved.\par
Next, let's prove $(4) \Rightarrow (1)$.
From Theorem \ref{mani} and Corollary 3,
there exists an element $ (2a,S) \in \tilde{{\cal U}}^*(g)$ 
such that $\Spur (2a,S) = 1$.
From \S \ref{random} and Theorem \ref{mani}, 
we have ${\cal D}_g^{\rho}(M^T_W)=1$.\par
Thus,
\begin{eqnarray*}
{\cal R}_g^{\rho}(2a,S;M^T_W) = 0.
\end{eqnarray*}
It implies that
\begin{eqnarray}
{\cal R}_g^{\rho}(2a,S;M^T(W^{-1}e_i ,J^{-1}e_i)) = 0. \Label{A6}
\end{eqnarray}
For $e \in \trp,~ \| e \| =1$,
\begin{eqnarray}
&~& R_g^{\rho}(2a,S ; x W^{-1} e ) \nonumber \\ 
&=& x^2 \rho - 2 x a (W^{-1} e) - S \nonumber \\
&=& ( x - J^{-1} a W^{-1} e ) \rho ( x - J^{-1} a W^{-1} e )
- S( a W^{-1} e ) 
-S \ge 0 . \Label{tan05}
\end{eqnarray}
Applying Lemma \ref{A2}, we obtain 
\begin{eqnarray}
- S(a W^{-1} e) -S = 0. \Label{tan1}
\end{eqnarray}
From (\ref{A6}), (\ref{tan05}) and Lemma \ref{A3}, substitution of
$e_i$ into $e$ implies that
\begin{eqnarray}
J a W^{-1} e = J^{-1} e.
\end{eqnarray}
Thus $a = W$. From (\ref{tan1}), we have 
\begin{eqnarray*}
S(e)= S,\quad  \forall e \in T_\rho P, \| e \| =1.
\end{eqnarray*}
Therefore, we get the condition $(1)$.
\end{pf}
\begin{lem}\Label{A2}
If Hermite matrixes $X,S$ and a density $\rho$
satisfy that
\begin{eqnarray*}
(x- X) \rho (x- X) + S \ge 0, \forall x \in \re,
\end{eqnarray*}
and that $\trh S= 0 $,
then $S=0$.
\end{lem}
\begin{pf}
Let 
$X= \sum_i x_i | \psi_i \rangle \langle \psi_i|$ be 
the spectral decomposition of $X$.
Since
\begin{eqnarray*}
\langle \psi_i | (x_i - X) \rho (x_i - X) + S|\psi_i \rangle
\ge 0,
\end{eqnarray*}
$\langle \psi_i | S |\psi_i \rangle\ge 0$.
Since $\trh S= 0 $,
$\langle \psi_i | S |\psi_i \rangle= 0$.
Next, we consider the following matrix
\begin{eqnarray*}
0 &\le& \left(
\begin{array}{cc}
\langle \psi_i | (x_i - X) \rho (x_i - X) + S|\psi_i \rangle &
\langle \psi_i | (x_i - X) \rho (x_i - X) + S|\psi_j \rangle \\
\langle \psi_j | (x_i - X) \rho (x_i - X) + S|\psi_i \rangle & 
\langle \psi_j | (x_i - X) \rho (x_i - X) + S|\psi_j \rangle
\end{array}
\right) \\
&=& 
\left(
\begin{array}{cc}
0 &
\langle \psi_i |  S|\psi_j \rangle \\
\langle \psi_j |  S|\psi_i \rangle & 
\langle \psi_j | (x_i - X) \rho (x_i - X) + S|\psi_j \rangle
\end{array}
\right).
\end{eqnarray*}
It implies that $\langle \psi_i |  S|\psi_j \rangle= 0$.
Therefore $S=0$.
\end{pf}
\begin{lem}\Label{A3}
Let $X,Y$ be Hermite matrixes. 
If 
\begin{eqnarray*}
\int_{\re} (x- X) \rho (x-X) M^T_Y(\,d x) = 0,
\end{eqnarray*}
then $X=Y$.
\end{lem}
It is easy.
\section{3-parameter Spin 1/2 model}\Label{quadra}
In this section, we will prove that if ${\cal H}={\bf C}^2$, then 
$\trp$ is random model.
Let us define the Pauli matrices $\sigma_{1},\sigma_{2},\sigma_{3}$ in the 
usual way:
\[
   \sigma_{1} = \left(
                 \begin{array}{cc}
                  0 & 1 \\
                  1 & 0 
                 \end{array}
                \right) 
                ,~ 
   \sigma_{2} = \left(
                 \begin{array}{cc}
                  0 & -i \\
                  i & 0 
                 \end{array}
                \right),~
   \sigma_{3} = \left(
                 \begin{array}{cc}
                  1 & 0 \\
                  0 & -1 
                 \end{array}
                \right) .
                \] 

Assume that $\trp = {\cal T}^0_{sa}({\bf C}^2)$,
$\rho= \frac{1}{2}(Id + \alpha \sigma_{3})$,$-1 \,< \alpha \,< 1$
and that $g$ is a quadratic form on $T_{\rho}P$.
 $f_{3} = \frac{\sqrt{1-\alpha^{2}}}{2} \sigma_{3},
~f_{i}= \frac{\sigma_{i}}{2}~,(i=1,2)$ are orthonormal bases on $\trp$.
The dual bases of $f^{i}$ are 
$f^3= \frac{-\alpha}{\sqrt{1-\alpha^2}} \id + \frac{1}{\sqrt{1-\alpha^{2}}} \sigma_3,~f^i=\sigma_i~(i=1,2)$ .
We need the following lemma.
\begin{lem}
If $e \in \trp ,~ \| e \| =1$, then
\begin{eqnarray}
  J^{-1} (e) \cdot \rho \cdot J^{-1} (e) =   \idh -\rho \Label{spi1}.
\end{eqnarray}
\end{lem}
\begin{pf}
We have 
\begin{eqnarray}
J^{-1}(e) = y^3 \frac{1}{1- \alpha^{2}}(- \alpha \idh + \sigma_{3}) + \sum_{i=2}^{3} e^{i} \sigma_{i}. 
\end{eqnarray}
Since there exists $t \in {\bf R}$ such that 
$\exp (\sqrt{-1} t \sigma_{3}) (e^1 \sigma_1 +e^2 \sigma_{2} )  
\exp (-\sqrt{-1} t \sigma_{3})
= \sqrt{(y^{1})^{2} + (y^{2})^{2}} \sigma_{1}$,
we may assume that $e^2=0$.
Then we have
\begin{eqnarray*}
 J^{-1} (e) \cdot \rho \cdot J^{-1} (e) 
&=& 
\left(
\begin{array}{cc}
\frac{-\alpha +1}{\sqrt{1-\alpha^{2}}}e^3 & e^1 \\
e^1 & \frac{-\alpha -1}{\sqrt{1-\alpha^{2}}}e^3
\end{array}
\right)
\left(
\begin{array}{cc}
\frac{1+\alpha}{2}& 0 \\
0 & \frac{1-\alpha}{2}
\end{array}
\right)
\left(
\begin{array}{cc}
\frac{-\alpha +1}{\sqrt{1-\alpha^{2}}}e^3 & e^1 \\
e^1 & \frac{-\alpha -1}{\sqrt{1-\alpha^{2}}}e^3
\end{array}
\right) \\
&=& 
\left(
\begin{array}{cc}
\frac{1-\alpha}{2} & 0 \\
0 & \frac{1+\alpha}{2} 
\end{array}
\right) = \idh - \rho.
\end{eqnarray*}
\end{pf}
We obtain the following theorem.
\begin{thm}
When ${\cal H}={\bf C}^2$,  
$\trp$ is a random model. 
\end{thm}
\section{Conclusions}
We have 
found a necessary and sufficient condition 
that a Cram\'{e}r-Rao type bound is attained by a random measurement.
But, we don't know the condition (1) or (2) in Theorem \ref{randomness1}
very well.We know no random model whose dimension is greater than 3.
Thus, it is conjectured that 
when $\trp$ is a random model, the dimension of $\trp$ is limited.
\par
\aki\noindent
{\Large\bf Acknowledgments}
\par\aku
I wish to thank Dr. A. Fujiwara for introducing me into this subject,
and Prof. K. Ueno for useful comments about this paper.
I also benefited from e-mail discussions with Dr. K. Matsumoto.
\newpage\noindent
{\LARGE\bf Appendices}
\appendix
\section{L-stable set}\Label{sec3}
The purpose of this section is proving the following theorem about a finite dimensional real vector space $W$ and its normal convex cone $L$.
\begin{defi}\rm
We assume that  $C \subset L$  is $L$-stable and convex.
A continuous map $Q:(L^{*})^i \to C $ is called {\it $C$-conic}
if
$\lll f,x \ggg \ge \lll f, Q(f) \ggg$ for arbitrary 
$x \in C ~,~f \in (L^{*})^i $,
where we denote $X^i$ the inner of a topological space $X$.
\end{defi}
\begin{thm}\Label{Lcqc}
Let $C$ be a subset of $L$. We assume that $C$ is $L$-stable and convex.
If there exists a $C$-conic map $Q$, then
\begin{eqnarray}
\im Q \subset K(C) \subset \overline{ \im Q } \Label{quasi20}
\end{eqnarray}
\end{thm}
\begin{lem}\Label{quasi7}
When $Q:(L^{*})^i \to L$ 
is continuous, then the following are equivalent.
\begin{eqnarray*}
&{\rm (1)}& \forall f \in (L^{*})^i~,~x \in \im Q~,~\lll f,x \ggg \ge \lll f,Q(f) \ggg \\
&{\rm (2)}& \forall f \in (L^{*})^i~,~x \in \im Q \setminus\{ Q(f) \} ~,~\lll f,x \ggg \,> \lll f,Q(f) \ggg 
\end{eqnarray*}
\end{lem}
\begin{pf}
(2) $\Rightarrow$ (1) is trivial. We prove that (1) $\Rightarrow$ (2).
Let $f, k \in (L^{*})^i$. 
It is sufficient to verify that if 
\begin{eqnarray}
\lll f, Q(k) \ggg  =\lll f,Q(f)\ggg ,\Label{quasi71} 
\end{eqnarray}
then $Q(f)=Q(k)$.\par
\noindent Step 1: We will prove $\lll k, Q(k)-Q(f) \ggg=0$.

Let $\alpha:=k-f$.
By the assumption of (1), for $1 \,> t \,> 0$,
\begin{eqnarray}
\lll t \alpha + f,  Q(t \alpha + f) -Q( \alpha + f) \ggg &\le& 0 \Label{quasi3}\\
\lll \alpha + f, Q(t \alpha  + f) -Q( \alpha + f) \ggg &\ge& 0 \Label{quasi4} \\
\lll f , Q(t \alpha + f ) - Q(f) \ggg &\ge& 0 . \Label{quasi41}
\end{eqnarray}
From (\ref{quasi3}) and (\ref{quasi4}),
\begin{eqnarray*}
\lll f, Q(t \alpha + f) -Q( \alpha + f) \ggg \le 0. \label{quasi1}
\end{eqnarray*}
Because of (\ref{quasi41}) and (\ref{quasi71}),
\begin{eqnarray}
\lll f, Q(t \alpha  + f) -Q(\alpha+f) \ggg \ge 0 . \Label{quasi42}
\end{eqnarray}
By (\ref{quasi41}) and (\ref{quasi42}),
\begin{eqnarray}
\lll f, Q(t \alpha  + f) -Q(\alpha+f) \ggg = 0 . \Label{quasi2}
\end{eqnarray}
By (\ref{quasi2}) and (\ref{quasi4}),
\begin{eqnarray}
\lll \alpha, Q(t \alpha  + f) -Q( \alpha + f) \ggg \ge 0 \Label{qua21}.
\end{eqnarray}
By (\ref{quasi2}) and (\ref{quasi3}),
\begin{eqnarray}
\lll \alpha, Q(t \alpha  + f) -Q( \alpha + f) \ggg \le 0 \Label{qua22}.
\end{eqnarray}
Because of (\ref{qua21}) and (\ref{qua22}),
\begin{eqnarray}
\lll \alpha, Q(t \alpha  + f) -Q( \alpha + f) \ggg = 0 \Label{qua23}.
\end{eqnarray}
From (\ref{qua23}) and (\ref{quasi71}),
\begin{eqnarray}
\lll \alpha+f,  Q(t \alpha  + f) -Q( \alpha + f) \ggg = 0 \Label{qua24}.
\end{eqnarray}
By the continuity of $Q$, we obtain 
\begin{eqnarray}
\lll k, Q(f) -Q(k) \ggg = 0 \Label{qua25}.
\end{eqnarray}
Step 2: Let $ h \in (L^{*})^i$ such that $h \neq f,k$.
We will prove $\lll h,Q(f)-Q(k)\ggg=0$.
By Step 1, we may assume that $\lll h,Q(f)-Q(k) \ggg \ge 0$.
Let $\beta:=h-f$, $1 \,> t \,> 0$. 
From the definition of $Q$,
\begin{eqnarray}
\lll t \beta  + f, Q(t \beta  + f) - Q(k) \ggg &\le& 0 \Label{quasi6} \\
\lll f, Q(t  \beta  + f) - Q(f) \ggg &\ge& 0. \Label{quasi5}
\end{eqnarray}
By (\ref{quasi71}) and (\ref{quasi5}),
$
\lll f, Q(t \beta + f) - Q(k) \ggg \ge 0 
$.
From (\ref{quasi6}) and the preceding inequality,
$
\lll \beta,Q(t \beta + f) - Q(k) \ggg \le 0 
$. 
By the continuity of $Q$,
$
\lll \beta,Q( f) - Q(k) \ggg \le 0 
$.
Therefore, we obtain
$
\lll h,Q( f) - Q(k) \ggg \le 0 
$. 
By the hypothesis,
$
\lll h,Q( f) - Q(k) \ggg \ge 0 
$.
Therefore $
\lll h,Q( f) - Q(k) \ggg = 0 
$.
We obtain $
Q( f) - Q(k)  = 0 
$. Therefore, the proof is complete.
\end{pf}
\begin{defi}\rm
A continuous map $Q:(L^{*})^i \to L$ 
is {\it quasi conic} if 
it satisfies the condition of lemma \ref{quasi7}.
\end{defi}
\begin{lem}\Label{quasi81}
Let $Q$ a quasi conic map.
When $C$ is  $L$ stable and convex and $\im Q \subset C$,the following are equivalent: \par
\begin{eqnarray*}
&{\rm (1)}& f \in (L^{*})^i,~x \in C ~\Rightarrow ~\lll f,x \ggg  \ge \lll f,Q(f) \ggg \\
&{\rm (2)}& f \in (L^{*})^i,~x \in C ,~
x \neq Q(f)~\Rightarrow ~\lll f,x \ggg \,> \lll f, Q(f) \ggg .
\end{eqnarray*}
\end{lem}
\begin{pf}
(2) $\Rightarrow$ (1) is trivial.
We will prove that (1) $\Rightarrow$ (2) by reductive absurdity.
There exist $f \in (L^{*})^i$ and $x \in C$
such  that  $x \neq Q(f),~
\lll f, Q(f) \ggg \ge \lll f,x \ggg$.
By the hypothesis, $\lll f,Q(f) \ggg=\lll f,x \ggg$.
$y,~f(\lambda)$ and $x(\lambda)$ are defined as follows:
for $\lambda \,>0$,
\begin{eqnarray} 
y:=x-Q(f),~ 
f(\lambda):=f-\lambda g(y),~
x(\lambda):=Q(f(\lambda)) - Q(f). \Label{qua36}
\end{eqnarray}
By the definition of $Q$, we obtain 
\begin{eqnarray}
\lll f(\lambda),x(\lambda) \ggg &\le& \lll f(\lambda),y \ggg \Label{qua34}\\
\lll f,x(\lambda) \ggg &\ge& \lll f,y \ggg =0 \Label{qua35}
\end{eqnarray}
From (\ref{qua34}) and (\ref{qua35}),
\begin{eqnarray+}
\lll f,x(\lambda) \ggg -\lambda \lll g(y),x(\lambda) \ggg 
&=& \lll f( \lambda) , x( \lambda) \ggg &\hbox{by}~(\ref{qua36}) \\
&\le& \lll f(\lambda),y \ggg ~&\hbox{by}~(\ref{qua34}) \\
&=& \lll f , y \ggg - \lambda \lll g(y) , y \ggg &\hbox{by}~(\ref{qua36}) \\
&=& - \lambda \lll g(y) , y \ggg &\hbox{by}~(\ref{qua35}) .
\end{eqnarray+}
Thus,
\begin{eqnarray}
\lambda \lll g (y), x(\lambda)- y \ggg \ge \lll f,x(\lambda)\ggg \,> 0 .
\end{eqnarray}
Hence, $\lll g(y), x(\lambda)-y \ggg \,> 0$.
Thus, $ \| y- x(\lambda)/2 \|^{2} \,< \| x(\lambda) \|^{2}/4$
i.e. $ \| y- x(\lambda)/2 \| \,< \| x(\lambda) \|/2$.
Therefore,
\begin{eqnarray}
\| x(\lambda) \| - \| y \| 
&\ge& \| x(\lambda) \| 
- ( \| y - \frac{x(\lambda)}{2} \| + \|\frac{x(\lambda)}{2} \| ) \nonumber \\
&=&  \|\frac{x(\lambda)}{2} \| -  \| y - \frac{x(\lambda)}{2} \| \nonumber \\
&\,>& 0 \Label{qua37}.
\end{eqnarray}
But by the continuity of $Q$,
$\lim_{\lambda\to 0}Q(f(\lambda)) = Q(f)$.
Thus $\lim_{\lambda \to 0} \| x(\lambda) \| =0$.
From (\ref{qua37}), $y=0$. We obtain a contradiction.
Therefore, we have (2). 
\end{pf}
\begin{lem}\Label{aha}
We obtain the following relations:
\begin{eqnarray}
B(C, (L^*)^i) \subset E(C,L) \subset \overline{B(C, (L^*)^i)},
\end{eqnarray}
where
\begin{eqnarray}
B(C, (L^*)^i) := \{ x \in C | 
\exists f \in (L^{*})^i ~,~\forall x \in C~,~f(x) \le f(y) \}.
\end{eqnarray}
\end{lem}
To know a proof of this lemma, see Ref 9.\par
\aku\noindent{\bf Proof of Theorem \ref{mainr} }\quad
From Lemma \ref{Lcqc} and Lemma \ref{quasi81}, 
If $Q$ is $C$-conic, then
$\im Q = B(C, (L^*)^i)$.
By Lemma \ref{aha}, we obtain (\ref{quasi20}). 
\leavevmode\hfill$\Box$\par

\section{Proof of Theorem \protect\ref{mani}}\Label{mainth}
It is the purpose of this section to prove the Theorem \ref{mani}.
Theorem \ref{mani} is described as follows.
\par\noindent{\bf Theorem \ref{mani}}\quad\it
We obtain 
\begin{eqnarray*}
 \inf_{M \in {\cal U}(T_{\rho}P)}{\cal D}^{\rho}_{g}(M) 
= \sup_{ (a,S) \in \tilde{{\cal U}}^{*}(g) } \Spur (a, S),
\end{eqnarray*}
where 
\begin{eqnarray*}
\tilde{{\cal U}}^{*}(g) 
&:=& \{ (a , S)\subset \End(T_{\rho}P) \times {\cal B}_{sa}^{*}({\cal H})  |
\forall x \in \trp ~,~R_g^{\rho}(a,S;x) 
\in {\cal B}_{sa}^{*,+}({\cal H}) \} \\
\Spur (a,S) &:=& 
\trt a + \lll S, \idh \ggg \\
R_g^{\rho}(a,S;x) &:=&
g(x,x) \cdot \rho - S - a(x). 
\end{eqnarray*}
\rm
The purpose of this section is to prove the preceding theorem by applying the following duality theorem.
\subsection{Infinite dimensional duality theorem (linear programming)}
Let ${\cal X},{\cal Y}$ be a locally convex Hausdorff real topological linear space,
 ${\cal A}$ a continuous linear operator form ${\cal X}$ to ${\cal Y}$
and ${\cal L}$ a closed convex cone in ${\cal X}$.
Let ${\cal X}^{*},{\cal Y}^{*}$ be the topological dual space of ${\cal X},{\cal Y}$
and ${\cal A}^{*}$ the continuous adjoint map of ${\cal A}$.
Let ${\cal L}^{*}$ be the conjugate cone of ${\cal L}$ in ${\cal X}^{*}$ 
i.e. ${\cal L}^{*}:=\{f \in {\cal X}^{*}|\forall x \in {\cal L}~,~f(x) \ge 0\}$. 
\begin{defi}\rm
Let ${\cal C}$ be an element of ${\cal X}^{*},
{\cal B}$ an element of ${\cal Y}$. 
We define ${\cal F}_{{\cal A},{\cal C}} ~,~{\cal E}_{{\cal B}}\subset {\bf R} \times {\cal Y}$ below:
\begin{eqnarray*}
 {\cal F}_{{\cal A},{\cal C}} &:=& \{(r,y) \in {\bf R} \times {\cal Y} |r=({\cal C},x),y={\cal A}x~ for~ some~ x \in {\cal L}\} \\
{\cal E}_{{\cal B}} &:=& {\bf R} \times \{ {\cal B} \}.
\end{eqnarray*}
\end{defi}
\begin{defi}\rm
Let ${\cal C}$ be an element of ${\cal X}^{*}$ 
and ${\cal B}$ an element of ${\cal Y}$.
$({\cal A},{\cal B},{\cal C})$ is called {\it normal}, if 
$\overline{{\cal F}_{{\cal A},{\cal C}} \cap {\cal E}_{{\cal B}}} =\overline{ {\cal F}_{{\cal A},{\cal C}}} \cap {\cal E}_{{\cal B}}$.
\end{defi}
\begin{thm}{[General duality theorem]\par}\Label{dual}
We obtain the following inequality for ${\cal C} \in {\cal X}^{*},{\cal B} \in {\cal Y}$.
We have the equality in (\ref{gdt}),
iff $({\cal A},{\cal B},{\cal C})$ is normal.
We assume that $\inf ({\cal C},x)= +\infty$, $\sup (f,{\cal B})= -\infty$
in the case of $\{x \in {\cal L} | {\cal A}x={\cal B} \}= \emptyset$,
$\{f \in {\cal Y}^{*}|{\cal C}-{\cal A}^{*}f \in {\cal L}^{*}\}= \emptyset$ 
with respectively.
\begin{eqnarray}
\inf_{\{x \in {\cal L} | {\cal A}x={\cal B} \}} \lll {\cal C},x \ggg \ge \sup_{\{f \in {\cal Y}^{*}|{\cal C}-{\cal A}^{*}f \in {\cal L}^{*}\}} \lll f,{\cal B} \ggg \Label{gdt}.
\end{eqnarray}
\end{thm} 
To know this theorem, see Ref. 9.
 To apply this theorem to the proof of Theorem \ref{mani},
we have to define ${\cal X},{\cal Y},{\cal L},{\cal A},{\cal B},{\cal C}$ 
such that $\{ x \in {\cal L} | {\cal A}x={\cal B} \}={\cal U}(T_{\rho}P)~,
~{\cal D}_{W}^{\rho}(M)=\lll {\cal C},M \ggg$.
\subsection{Topology}
To apply Theorem \ref{dual} to the proof of Theorem \ref{mani}
we will construct ${\cal X},{\cal L}$. 
\begin{defi}\rm
${\cal X}\bigl(T_{\rho}P,{\cal H},g \bigr)$ is defined the set of the map 
$M:{\cal B}(T_{\rho}P) \to {\cal B}_{sa}({\cal H} )$
which satisfies the following conditions:
\begin{eqnarray}
 &\circ & ~~M \Bigl(\Cup_{\lambda \in \Lambda}B_{\lambda} \Bigr)
=\sum_{\lambda \in \Lambda}M(B_{\lambda}) ~~~(
B_{\lambda} \in {\cal B}(T_{\rho}P)
~,~\lambda_1 \neq \lambda_2 \in \Lambda~\Rightarrow~
B_{\lambda_1} \cap B_{\lambda_2} = \emptyset
~,~| \Lambda | = \aleph_{0}) \nonumber \\
 &\circ& ~~\sup_{B \in {\cal B}(T_{\rho}P)} \| M(B) \| \,< \infty 
\Label{yukaia} \\
 &\circ& ~~\forall f \in T_{\rho}^{*}P~,~\forall y \in T_{\rho}P ~,~
\Bigl| \intt f(x) \trh M(\,dx) y \Bigr| \,< \infty \Label{yukaib} \\
 &\circ & ~~\Bigl| \intt g(x,x) \trh M(\,dx) \rho \Bigr| \,< \infty \label{yukaid}.
\end{eqnarray}
\end{defi}
As ${\cal B}_{sa}({\cal H})$ is a vector space,  ${\cal X}\bigl(T_{\rho}P,{\cal H},g \bigr)$ is a vector space, too.\par
The norm $\| \cdot \|$ of $\End(T_{\rho}P)$ is defined in the following:
\begin{eqnarray}
\| A \| := \sqrt{\trt A A^{*} },~\forall A \in \End(T_{\rho}P).
\end{eqnarray}
The topology of $\End(T_{\rho}P) $ is defined by this norm.
The map $E:{\cal X}(T_{\rho}P,{\cal H},g) \to \End(T_{\rho}P) $ is defined 
in the following:
\begin{eqnarray}
\begin{array}{cccc}
 E(M):& T_{\rho}P & \to & T_{\rho}P \\
 &\tin& &\tin\\
 &y&\mapsto&\intt x \trh M(\,dx) y
\end{array}
,~\forall M \in {\cal X}(T_{\rho}P,{\cal H},g).
\end{eqnarray}
This definition of $E$ is well defined by condition (\ref{yukaib}).
\par 
We will define the topology of ${\cal X}\bigl(T_{\rho}P ,{\cal H},g \bigr)$.
For this definition, a norm $\| \cdot \|_1$ and 
two semi-norms $\| \cdot \|_2, \| \cdot \|_3$ 
in ${\cal X}\bigl(T_{\rho}P ,{\cal H},g \bigr)$ are defined as follows:
for $ M \in {\cal X}(T_{\rho}P,{\cal H},W)$,
\begin{eqnarray+}
\| M \|_{1} &:=& \sup_{B \in {\cal B}(T_{\rho}P)} \| M(B) \|
& {\it by } (\ref{yukaia}) \\
\| M \|_{2} &:=& \| E(M) \|
& {\it by } (\ref{yukaib}) \\
\| M \|_{3} &:=& \Bigl|  \intt g(x,x) \trh M(\,dx) \rho \Bigr|
& {\it by } (\ref{yukaid}) .\\
\end{eqnarray+}
A norm $\| \cdot \|$ in ${\cal X}(T_{\rho}P,{\cal H},g)$
is defined in the following:
\begin{eqnarray}
\| M \| := \|M \|_{1}+ \|M\|_{2} +\|M\|_{3} ,
~ \forall M \in {\cal X}(T_{\rho}P,{\cal H},W) .
\end{eqnarray}
We define the topology of ${\cal X}(T_{\rho}P,{\cal H},g)$ by this norm.
A closed convex cone ${\cal L}(T_{\rho}P,{\cal H},g)$ in 
 ${\cal X}\bigl(T_{\rho}P,{\cal H},g \bigr)$ is defined as follows:
\begin{eqnarray}
{\cal L}(T_{\rho}P,{\cal H},g):= \{ M \in {\cal X}(T_{\rho}P,{\cal H},g) |
\forall B \in {\cal B}(T_{\rho}P)~,~ M(B) \in {\cal B}_{sa}^{+}({\cal H}) \}.
\end{eqnarray}
\begin{lem}
${\cal L}(T_{\rho}P,{\cal H},g)$ is a closed convex cone.
\end{lem}
\begin{pf}
It is trivial that it is a convex cone.
We have to prove that it is a closed set.
Let $\{ M_{k} \}$ be a converging sequence of ${\cal L}(T_{\rho}P,{\cal H},g)$.
Its convergence point is denoted by
$M \in {\cal X}(T_{\rho}P,{\cal H},g)$.
It suffices to prove that 
$M$ is included in ${\cal L}(T_{\rho}P,{\cal H},W)$.
Since $\|M_{k} - M\| \to 0$, then $\| M_{k} - M \|_{1} \to 0$.
Because $\|M_{k}(B)-M(B)\| \to 0$, $M_{K}(B) \in {\cal B}_{sa}^{+}({\cal H})$,
and ${\cal B}_{sa}^{+}({\cal H})$ is a closed convex cone,
we obtain $M(B) \in {\cal B}_{sa}^{+}({\cal H})$.
Therefore, $M \in {\cal L}(T_{\rho}P,{\cal H},g)$.
\end{pf}
\begin{lem}
The map $E :{\cal X}(T_{\rho}P,{\cal H},g) \to \End(T_{\rho}P ) $ is a continuous linear map.
\end{lem}
\begin{pf}
The linearity is trivial.
We will prove the map is bounded.
\begin{eqnarray*}
\| E(M) \| = \| M \|_{2} \le \| M \| .
\end{eqnarray*}
\end{pf}
\begin{defi}\rm
The map $\Int$ from ${\cal X}(T_{\rho}P,{\cal H},g)$ to ${\cal B}_{sa}({\cal H})$
is defined in the following:
\begin{eqnarray}
\Int(M) :=  M(T_{\rho}P) ,~\forall M \in {\cal X}(T_{\rho}P,{\cal H},g).
\end{eqnarray}
\end{defi}
\begin{lem}
The map $\Int$ is a continuous linear map.
\end{lem}
\begin{pf}
The linearity is trivial.
We prove that the map is bounded.
\begin{eqnarray*}
\| \Int(M) \| = \| M(T_{\rho}P) \| \le \| M \|_{1} \le \| M \|.
\end{eqnarray*}
Thus, it is bounded.
\end{pf}
\begin{defi}\rm
The map $C:{\cal X}(T_{\rho}P,{\cal H},g) \to {\bf R}$ is defined as follows:
\begin{eqnarray}
C(M) := \intt g(x,x) \trh M(\,dx) \rho
,~\forall M \in {\cal X}(T_{\rho}P,{\cal H},g) .
\end{eqnarray}
\end{defi}
\begin{lem}
$C$ is a bounded linear functional.
\end{lem}
\begin{pf}
The linearity is trivial.
\begin{eqnarray*}
\|C(M) \| = \| M \|_{3} \le \|M \|.
\end{eqnarray*}
Thus, it is bounded.
\end{pf}
An element $M$ of ${\cal L}(T_{\rho}P,{\cal H},g)$ is an element of ${\cal U}(T_{\rho}P)$, iff
\begin{eqnarray}
E(M)=\idt ~,~\Int(M)=\idh.
\end{eqnarray}
For $M \in {\cal U}(T_{\rho}P)$,
\begin{eqnarray}
{\cal D}_{g}^{\rho}(M) = C(M).
\end{eqnarray}
\subsection{Applying the infinite linear programming duality theorem}
We put in the following:
\begin{eqnarray*}
{\cal X} &:=& {\cal X}(T_{\rho}P,{\cal H},g) \\
{\cal Y} &:=& \End(T_{\rho}P) \times {\cal B}_{h}({\cal H}) \\
{\cal L} &:=& {\cal L}(T_{\rho}P,{\cal H},g) \\
{\cal A} &:=& E \times \Int \\
{\cal B} &:=& (\idt , \idh) \\
{\cal C} &:=& C .
\end{eqnarray*}
From the preceding discussion ${\cal X},{\cal Y},{\cal L},{\cal A},{\cal B},{\cal C}$ satisfy the condition of Theorem \ref{dual}.
Thus,
\begin{equation}
 \inf_{M \in {\cal U}(T_{\rho}P)}{\cal D}^{\rho}_{g}(M)  
= \inf_{\{x \in {\cal L} | {\cal A}x={\cal B} \}} \lll {\cal C},x \ggg \Label{inf}.
\end{equation}
Therefore, to prove Theorem \ref{mani},
we have to prove the following equation:
\begin{eqnarray}
 \sup_{ (a,S) \in {\cal U}^{*}(g) } \Spur (a,S) 
= \sup_{\{f \in {\cal Y}^{*}|{\cal C}-{\cal A}^{*}f \in {\cal L}^{*}\}} \lll f,{\cal B} \ggg .
\end{eqnarray}
Notice that ${\cal Y}^{*}= \End(T_{\rho}P) \times {\cal B}_{sa}^{*}({\cal H})$.
$\End(T_{\rho}P)$ is regarded as the dual space of itself by
\begin{eqnarray}
\begin{array}{cccc}
\lll~,~\ggg:& \End(T_{\rho}P) \times \End(T_{\rho}P) & \to &{\bf R} \\
& \tin && \tin \\
& (A,B) & \mapsto & \lll A, B \ggg =\trt AB .
\end{array}
\end{eqnarray}
\begin{lem}\Label{sup}
For $(a,S) \in \End(T_{\rho}P) \times {\cal B}_{sa}^{*}({\cal H}) $, the following are equivalent.
\begin{eqnarray}
&\circ& ~~(a,S) \in {\cal U}(g) \\
&\circ& ~~{\cal C} - {\cal A}^{*}(a,S) \in {\cal L}^{*}.
\end{eqnarray}
\end{lem}
From this Lemma, we obtain 
\begin{eqnarray}
 \sup_{ (a,S) \in {\cal U}^{*}(g) } \Spur (a,S)
= \sup_{\{f \in {\cal Y}^{*}|{\cal C}-{\cal A}^{*}f \in {\cal L}^{*}\}} \lll f,{\cal B} \ggg . \Label{supb}
\end{eqnarray}
Thus, if it is proved that $({\cal A},{\cal B},{\cal C})$ is  normal,
the proof of Theorem \ref{mani} is complete.

\begin{pf}
\begin{eqnarray}
{\cal C} - {\cal A}^{*}(a,S) = {\cal C} - E^{*}(a) - \Int^{*}(S) .
\end{eqnarray}
For $x \in T_{\rho}P ~,~ P \in {\cal B}_{sa}^{+}({\cal H})$
$M_{P,x} \in {\cal L} \bigl(T_{\rho}P ,{\cal H},g \bigr)$ 
is defined in the following:
\[ M_{P,x}(B):=\left\{
\begin{array}{@{\,}ll}
 0 & (x \notin B  )\\
 P & (x \in B ) 
\end{array}
\right. . \]
Thus, the following are equivalent.
\begin{eqnarray}
&\circ&~~ {\cal C} - E^{*}(a) - \Int^{*}(S) \in {\cal L}^{*} \\
&\circ&~~ \forall x \in T_{\rho}P~,~\forall P \in {\cal B}_{sa}({\cal H})~,~\lll {\cal C} - E^{*}(A) - \Int^{*}(S) ,M_{P,x} \ggg \ge 0 .
\end{eqnarray}
Therefore,
\begin{eqnarray}
\lll {\cal C},M_{P,x} \ggg 
&=& \intt g(y,y) \trh M_{P,x}(\,dy) \rho \nonumber \\
&=& g(x,x) \lll \rho,P \ggg .
\end{eqnarray}
Let the map $U:T_{\rho}P \to {\cal T}_{sa}({\cal H})$ be a trivial embedding.
As ${\cal T}_{sa}^{*}({\cal H}) = {\cal B}_{sa}({\cal H})$ 
with respect to the norm topology,
we define $U^{*}:{\cal B}_{sa}({\cal H}) \to T_{\rho}^{*}P$ in the natural sense.
\begin{eqnarray}
\lll E^{*}(a), M_{P,x} \ggg
&=& \lll a , E(M_{P,x}) \ggg \nonumber \\
&=& \lll a ,  x \otimes U^{*}(P) \ggg ~~~(\hbox{From } \End(T_{\rho}P) \cong T_{\rho}P \otimes T_{\rho}^{*}P  )\nonumber \\
&=& \trt a \bigl(x \otimes U^{*}(P)\bigr) \nonumber  \\
&=& \trt a (x) \otimes U^{*}(P) \nonumber  \\
&=& \lll U^{*}P, a(x) \ggg \nonumber \\
&=& \lll P, a(x) \ggg \\
\lll \Int^{*}(S), M_{P,x} \ggg
&=& \lll S , \Int(M_{P,x}) \ggg \nonumber \\
&=& \lll S , P \ggg .
\end{eqnarray}
Therefore, we obtain
\begin{eqnarray}
 \lll {\cal C} - E^{*}(a) - \Int^{*}(S) ,M_{P,x} \ggg 
= \lll g(x,x) \rho - a(x) - S , P \ggg .
\end{eqnarray}
Thus, the following are equivalent.
\begin{eqnarray}
&\circ&~~ \forall P \in {\cal B}_{sa}^+({\cal H})~,~\lll {\cal C} - E^{*}(a) - \Int^{*}(S) ,M_{P,x} \ggg \ge 0 \\
&\circ&~~ \forall x \in \trp~,~ g(x,x) \rho - a(x) - S  \in {\cal B}_{sa}^{*,+}({\cal H}).
\end{eqnarray}
Therefore, the following are equivalent.
\begin{eqnarray}
&\circ&~~ {\cal C} - E^{*}(a) - \Int^{*}(S) \in {\cal L}^{*} \\
&\circ&~~\forall x \in T_{\rho}P~,~  g(x,x) \rho - a(x) 
- S  \in {\cal B}_{sa}^{*,+}({\cal H}).
\end{eqnarray}
Thus, the proof is complete.
\end{pf}
\subsection{Normality}
In this section, we prove that 
$({\cal A},{\cal B},{\cal C})$ is normal.
\begin{defi}\rm
The subsets ${\cal F~,~G~,~E}$ of ${\cal Y}$ are defined  in the following:
\begin{eqnarray*} 
 {\cal F} &:=& {\cal F}_{{\cal A},{\cal C}} 
  = \{ {\cal C}(M) \times {\cal A}(M) | M \in {\cal L}(T_{\rho}P,{\cal H},W) \} \\
 {\cal G} &:=& {\bf R} \times \End(T_{\rho}P) \times \{ \idh \} \\
 {\cal E} &:=& {\cal E}_{{\cal B}} 
=  {\bf R} \times \{(\idt , \idh ) \} .
\end{eqnarray*}
Notice that ${\cal G}$ and ${\cal E}$ are closed sets.
\end{defi}
\begin{lem}
If 
\begin{eqnarray}
\overline{{\cal F}} \cap {\cal G} = \overline{{\cal F} \cap {\cal G} }~,~
\overline{{\cal F} \cap {\cal G}} \cap {\cal E} = \overline{{\cal F} \cap {\cal E}}  ,
\end{eqnarray}
then
\begin{eqnarray}
\overline{{\cal F}} \cap {\cal E} = \overline{{\cal F} \cap {\cal E} } . 
\Label{set}
\end{eqnarray}
\end{lem}
\begin{pf}
\begin{eqnarray*}
\hbox{\rm The left-hand in (\ref{set})} = \overline{{\cal F}} \cap {\cal G} \cap {\cal E}   
 = \overline{{\cal F} \cap {\cal G}} \cap {\cal E}  
 = \overline{{\cal F} \cap {\cal E}} .
\end{eqnarray*}
\end{pf}
\begin{lem}
We obtain 
\[ \overline{{\cal F} \cap {\cal G}} \cap {\cal E} = \overline{{\cal F} \cap {\cal E}} . \]
\end{lem}
\begin{pf}
Let $e^{1}, \ldots , e^{n}$ be bases of $\trp$, and let 
$e_{1} , \ldots , e_{n}$ the dual bases of $e^1 , \ldots , e^n$.
Linear functionals $e_{i} \star e^{j}$, $g \star \rho$
on ${\cal X}(T_{\rho}P,{\cal H},g)$ 
are defined in the following way:
\begin{eqnarray*}
\begin{array}{cccc}
e_i \star e^j :&{\cal X}(T_{\rho}P,{\cal H},g) &\to& {\bf R} \\
&\tin&&\tin \\
& M &\mapsto& \intt \lll e_{i},x \ggg \trh M(\,dx) e^{j} \\ 
g \star \rho : &{\cal X}(T_{\rho}P,{\cal H},g) &\to& {\bf R} \\
&\tin&&\tin \\
&M &\mapsto& \intt g(x) \trh M(\,dx) \rho .
\end{array}
\end{eqnarray*}
As $e^{1}, \ldots ,e^{n}$ are linearly dependent,
let $D_i^+,D_i^-$ be nonnegative bounded selfadjoint operators such that 
$\lll e_{i}, e^j \ggg =\trh (D_{i}^{+} - D_{i}^{-}) e^j $ 
for $(1 \le j \le n)$. 
Therefore, we define that 
$D:=\sum_{i=1}^{n} (D_{i}^{+} +D_{i}^{-})\in {\cal B}_{sa}^+({\cal H})
,d:=\| D \|_{{\cal B}_{sa}({\cal H})} = \sup_{\{\phi \in {\cal H} | \|\phi \| =1\} } \| D \phi \|$.
\par
It suffices to prove that for any Cauchy sequence 
$\{ a_{k} \} \subset  {\cal F} \cap {\cal G}$ such that
$\lim_{k \to \infty} a_{k} \in {\cal E}$,
there exists a Cauchy sequence $\{ b_{k} \} \subset {\cal F} \cap {\cal E}$
 such that 
$\lim_{k \to \infty} a_{k} = \lim_{k \to \infty} b_{k} \in {\cal E}$.
The components of $a_{k}$ are denoted by
$a_{k}=d_{k} \times a_{k,j}^{~i} \times \idh \in {\bf R} \times \End(T_{\rho}P) \times {\cal B}_{sa}({\cal H})$.
Notice that
$E(M)_{j}^{i} = \lll e_{j} \star e^{i} , M \ggg $ 
for $M \in {\cal M}\bigl(T_{\rho}P ,{\cal H},g \bigr)$.
$c_{k}$ denotes the maximum $\max_{0 \le i \le n}\sum_{j=1}^{n} | a_{k,j}^{~i} - \delta_{j}^{i} |^{2}$.
Since there exists the limit of a sequence $a_{k}$, then we have
 $\lim_{k \to \infty} c_{k} =0$ i.e.
\[ \forall m \in {\bf N}~,~\exists k(m) \in {\bf N}~s.t.~c_{k(m)} \,< (md)^{-1} .\]
For $M_{k} \in ({\cal C} \times {\cal A} )^{-1}(a_{k}) \subset {\cal M}\bigl(T_{\rho}P ,{\cal H},g \bigr)$
elements $\{M_{m,1}~,~ M_{m,2} \}_{m=1}^{\infty} \in {\cal L} \bigl(T_{\rho}P ,{\cal H},g \bigr)$ are defined in the following: \par
\begin{eqnarray*}
 M_{m,1}(B) &:=& \frac{m-1}{m} \cdot M_{k(m)}(\frac{m-1}{m} \cdot B) ~,~~
\hbox{ for } \forall B \in {\cal B}({\bf R}^{n \times n}) \\
 M_{m,2} &:=& \sum_{i=1}^{n} \bigl( \delta_{m \cdot d \cdot \sum_{j=1}^{n}(\delta_{j}^{i} - a_{k(m),j}^{i})e^{j}} \cdot \frac{1}{m \cdot d} \cdot D_{i}^{+} + \delta_{m \cdot d \cdot \sum_{j=1}^{n}(- \delta_{j}^{i} + a_{k(m),j}^{~i})e^{j}} \cdot \frac{1}{m \cdot d} \cdot D_{i}^{-} \bigr), 
\end{eqnarray*}
where $\delta_{\sum_{j=1}^{n}a_{j}e^{j}}$ is the delta measure which 
takes value only $e_{j} (x) =a_{j}$ and for $c \in {\bf R}^+,
~B \in {\cal B}(\trp)$ the set $c \cdot B \in {\cal B}(\trp)$
 is defined as follows:
\[ c \cdot B := \{ x \in T_{\rho}P | c \cdot x \in B \} .\]
Thus,
\[ M_{m,1}(T_{\rho}P)= \frac{m-1}{m} \idh~,~M_{m,2}(T_{\rho}P) \le \frac{1}{m \cdot d} D \le \frac{1}{m} \idh . \] 
A measurement $M_{b,m}$ is defined in the following way:
\[ M_{b,m} := M_{m,1} + M_{m,2} + \delta_{0} \bigl( \frac{1}{m}\idh-M_{m,2}(T_{\rho}P) \bigr)  \in {\cal M}\bigl(T_{\rho}P ,{\cal H},g \bigr).\]
Thus,
\begin{eqnarray*}
E(M_{m,1})_j^i
=\lll e_{j} \star e^{i} , M_{m,1} \ggg 
= \lll e_{j} \star e^{i} , M_{k(m)} \ggg 
= a_{k(m),j}^{i}.
\end{eqnarray*}
Therefore,
\begin{eqnarray*}
&~& \int_{T_{\rho}P} e_{j}(x) M_{m,2}(\,d x) \\
 &=& \int_{T_{\rho}P} x_{j} M_{m,2}(\,d x) \\
 &=& \sum_{i=1}^{n} \bigl( m \cdot d \cdot (\delta_{j}^{i} - a_{k(m),j}^{~i}) \cdot \frac{1}{m \cdot d} \cdot D_{i}^{+} + m \cdot d \cdot (- \delta_{j}^{i} + a_{k(m),j}^{~i}) \cdot \frac{1}{m \cdot d} \cdot D_{i}^{-} \bigr) \\
 &=& \sum_{i=1}^{n} \bigl((\delta_{j}^{i} - a_{k(m),j}^{~i}) \cdot D_{i}^{+} + (- \delta_{j}^{i} + a_{k(m),j}^{~i}) \cdot D_{i}^{-} \bigr) \\
  &=& \sum_{i=1}^{n} (\delta_{j}^{i} - a_{k(m),j}^{~i}) \cdot (D_{i}^{+} - D_{i}^{-} ) .
 \end{eqnarray*}
Thus,
\begin{eqnarray*}
\lll e_{j} \star e^{i} , M_{m,2} \ggg 
&=& \trh ( \sum_{l=1}^{n} (\delta_{j}^{l} - a_{k(m),j}^{~l}) \cdot (D_{i}^{+} - D_{i}^{-} ) e^i) \\
&=& \delta_{j}^{i} - a_{k(m),j}^{~i} .
\end{eqnarray*}
Therefore,
\begin{eqnarray*}
\lll e_{j} \star e^{i} ,  M_{b,m} \ggg 
&=& \lll e_{j} \star e^{i} , M_{m,1}+M_{m,2} \ggg \\ 
&=& a_{k(m),j}^{~i} + \delta_{j}^{i} - a_{k(m),j}^{~i} \\
&=& \delta_{j}^{i} .
\end{eqnarray*}
$b_{m}$ denotes ${\cal C} \times {\cal A} (M_{b,m}) \in {\cal E}$.
Then it suffices to prove that $\lim_{ m \to \infty} b_{m} = \lim_{m \to \infty}a_{k(m)}$.
\begin{eqnarray*}
&~& \lll g \star \rho , M_{m,2} \ggg \\
 &=& \sum_{i=1}^{n} g(m \cdot d \cdot \sum_{j=1}^{n}(\delta_{j}^{i} - a_{k(m),j}^{~i})e^{j}) 
\cdot \trh (\frac{1}{m \cdot d} \cdot D_{i}^{+}\rho) \\
&~& + \sum_{i=1}^{n} g( m \cdot d \cdot \sum_{j=1}^{n}(- \delta_{j}^{i} + a_{k(m),j}^{~i})e^{j} ) 
\cdot \trh  (\frac{1}{m \cdot d} \cdot D_{i}^{-}\rho ) \\
 &=& (\max_{|x|=1}g(x)) \cdot \frac{1}{m \cdot d} 
\cdot \trh ( (\sum_{i=1}^{n} D_{i}^{+}+D_{i}^{-}) \rho) \\
 & \to & 0 ~({\it as}~m \to \infty).
\end{eqnarray*}
And
\begin{eqnarray*}
&~& \lll g \star \rho , M_{m,1} \ggg \\
&=& \int_{T_{\rho}P} g(x) \trh (M_{m,1}(\,d x) \rho) \\
&=& \int_{T_{\rho}P} \frac{m-1}{m}g(\frac{m}{m-1} \cdot x) \trh (M_{k(m)}(\,d x)\rho). 
\end{eqnarray*}
And
\begin{eqnarray*}
&~& \int_{T_{\rho}P} \frac{m-1}{m} g(\frac{m}{m-1} \cdot x) \trh (M_{k(m)}(\,d x)\rho) \\
&=& \int_{T_{\rho}P} \frac{m}{m-1} g(x) \trh (M_{k(m)}(\,d x)\rho) \\
&=& \frac{m}{m-1} \cdot \lll g \star \rho , M_{k(m)} \ggg . 
\end{eqnarray*}
Thus,
\[ \lll g \star \rho , M_{m,1} \ggg 
= \frac{m}{m-1} \cdot \lll g \star \rho , M_{k(m)} \ggg \]
As $ \{ \lll g \star \rho , M_{k(m)} \ggg \} $ is a Cauchy sequence,
\[ \lll g \star \rho ,M_{m,1} \ggg - \lll g \star \rho , M_{k(m)} \ggg \to 0 
~(~\hbox{as}~m \to \infty) .\]
We obtain that $\lim_{ m \to \infty} b_{m} = \lim_{m \to \infty}a_{k(m)}$.
The proof is complete.
\end{pf}
\begin{lem}
We obtain 
\[ \overline{{\cal F}} \cap {\cal G} = \overline{{\cal F} \cap {\cal G}} .\]
\end{lem}
\begin{pf}
It suffices to prove that 
for a Cauchy sequence $ \{ a_k  \} \subset  {\cal F} $ 
such that $\lim_{k \to \infty} a_{k} \in {\cal G}$ 
there exists a Cauchy sequence $\{ b_{k} \} \subset {\cal F} \cap {\cal G}$ 
such that $\lim_{k \to \infty} a_{k} = \lim_{k \to \infty} b_{k} \in {\cal G}$.
The component of $a_{k}$ is denoted by
$a_{k}=d_{k} \times a_{k,j}^{~i} \times X_{k} \in 
{\bf R} \times \End(T_{\rho}P) \times {\cal B}_{sa}({\cal H})$.
For $M_{k} \in ({\cal C} \times {\cal A})^{-1}(a_{k}) \subset {\cal M}\bigl(T_{\rho}P ,{\cal H},g \bigr)$, 
$M_{k,1}(B)$ is defined in the following:
\begin{eqnarray*}
 M_{k,1}(B) :=
\left \{
\begin{array}{@{\,}lll}
& M_{k}(B) & ( \| X_{k} \|_{{\cal B}_{sa}({\cal H})} \le 1 ) \\
&\frac{1}{\| X_{k} \|_{{\cal B}_{sa}({\cal H})}} \cdot M_{k}(\frac{1}{\| X_{k} \|_{{\cal B}_{sa}({\cal H})}} \cdot B)& 
(\| X_{k} \|_{{\cal B}_{sa}({\cal H})} \,> 1 )
\end{array}
\right.
~\hbox{for}~B \in {\cal B}(T_{\rho}P) .
\end{eqnarray*}
Notice that $M_{k}(T_{\rho}P) = X_{k} $.
$M_{b,k}$ is defined in the following way:
\[ M_{b,k} := M_{k,1} + \delta_{0} (\idh-M_{k,1}(T_{\rho}P)) . \]
If $\| X_{k} \|_{{\cal B}_{sa}({\cal H})} \,> 1$, then
\begin{eqnarray*}
\lll e_{j} \star e^{i} ,M_{b,k} \ggg &=& \lll e_{j} \star e^{i} ,  M_{k,1} \ggg \\ 
&=& \int_{T_{\rho}P} x_{j} \trh (M_{b,k}(\,d x) e^{i}) \\
&=& \int_{T_{\rho}P} \| X_{k} \|_{{\cal B}_{sa}({\cal H})} \cdot x_{j} 
\trh (\frac{1}{\| X_{k} \|_{{\cal B}_{sa}({\cal H})}} \cdot M_{k}(\,d x) e^{i}) \\
&=& \int_{T_{\rho}P} x_{j} \trh (M_{k}(\,d x) e^{i}) \\
&=& \lll e_{j} \star e^{i} ,M_{k} \ggg .
\end{eqnarray*}
If $\| X_{k} \|_{{\cal B}_{sa}({\cal H})} \le 1$, then
\begin{eqnarray*}
\lll e_{j} \star e^{i} , M_{b,k} \ggg = \lll e_{j} \star e^{i} , M_{k,1} \ggg 
= \lll e_{j} \star e^{i}, M_{k} \ggg .
\end{eqnarray*}
Thus,
\begin{eqnarray}
\lll e_{j} \star e^{i} , M_{b,k} \ggg = \lll e_{j} \star e^{i} , M_{k} \ggg .
\end{eqnarray}
If $\| X_{k} \|_{{\cal B}_{sa}({\cal H})} \,> 1$, then
\begin{eqnarray*}
\lll g \star \rho ,M_{b,k} \ggg &=& \lll g \star \rho , M_{k,1} \ggg \\
 &=& \int_{T_{\rho}P} g(x) \trh (M_{b,k}(\,d x) \rho) \\
&=& \int_{T_{\rho}P} g(\| X_{k} \|_{{\cal B}_{sa}({\cal H})} \cdot x) 
\trh (\frac{1}{\| X_{k} \|_{{\cal B}_{sa}({\cal H})}} \cdot M_{k}(\,d x) \rho) \\
&=& \int_{T_{\rho}P} \| X_{k} \|_{{\cal B}_{sa}({\cal H})} \cdot g(x) 
\trh (M_{k}(\,d x) \rho) \\
&=& \| X_{k} \|_{{\cal B}_{sa}({\cal H})} \lll g \star \rho , M_{k} \ggg .
\end{eqnarray*}
Thus,
\[ \lll g \star \rho ,  M_{b,k} \ggg 
= \| X_{k} \|_{{\cal B}_{sa}({\cal H})} \lll g \star \rho, M_{k} \ggg . \]
If $\| X_{k} \|_{{\cal B}_{sa}({\cal H})} \le 1$, then
\begin{eqnarray*}
\lll g \star \rho ,M_{b,k} \ggg = \lll g \star \rho ,  M_{k,1} \ggg
= \lll g \star \rho , M_{k} \ggg .
\end{eqnarray*}
Because  $\| X_{k} \|_{{\cal B}_{sa}({\cal H})} \to 1$ 
and $\{ \lll g \star \rho , M_{k} \ggg \}$ is a Cauchy sequence ,
\begin{eqnarray*}
 \lim_{k \to \infty} \lll g \star \rho , M_{b,k} \ggg 
= \lim_{k \to \infty} \lll g \star \rho ,  M_{k} \ggg . 
\end{eqnarray*}
Let $b_{k}:=({\cal C} \times {\cal A}) (M_{b,k})$, then
$\lim_{k \to \infty} a_{k} =\lim_{k \to \infty} b_{k}$.
The proof is complete.
\end{pf}
From the preceding lemmas, we obtain the following theorem.
\begin{thm}
$({\cal A},{\cal B},{\cal C})$ is normal.
\end{thm}
We obtain Theorem \ref{mani} from this theorem, Theorem \ref{dual}, 
the equation (\ref{inf}) and the equation (\ref{supb}).
\par\aki
\noindent{\Large\bf References}
\par\aki
$~^1$ A. S. Holevo, 
 \it Probabilistic and Statistical Aspects of Quantum Theory 
 \rm(North\_Holland, Amsterdam, 1982).
\par
$~^2$ H. P. Yuen and M. Lax, 
IEEE trans. {\bf IT-19}, 740 (1973).
\par
$~^3$ H. Nagaoka, 
Trans. Jap. Soci. Ind. App. Math.,
{\bf 1}, 305 
 (1991)(in Japanese).
\par
$~^4$ K. Matsumoto,
 ``A new approach to the Cram\'{e}r-Rao type bound of the pure state model,'' 
METR 96-09,(1996).
\par
$~^5$ R. M. Van Style and R. J. B. Wets, 
J. Math. Anal. Appl.,
{\bf 22}, 679 (1968).
\par
$~^6$ A. S. Holevo,
Rep. Math. Phys.,
{\bf 12}, 251 (1977).
\par
$~^7$ M. Hayashi,
Masters Thesis,
Dep. Math., Kyoto University,
(1996)(in Japanese).
\par
$~^8$ H. P. Yuen, M. Lax and R. S. Kennedy,
IEEE, {\bf IT-21}, 125 (1975).
\par
$~^9$ R. Hartley
Siam J. Appl. Math. {\bf 34}, 211 (1978).
\par
$~^{10}$ A. Fujiwara and H. Nagaoka,
in {\em Quantum coherence and decoherence},
edited by K. Fujikawa and Y. A. Ono,
(Elsevier, Amsterdam, 1996), pp. 303.
\par
$~^{11}$ A. Fujiwara and H. Nagaoka,
Phys. Lett. A 201 (1995) 119-124.
\end{document}